\documentclass[11pt,a4paper]{article}
\pdfoutput=1
\usepackage{jheppub}

\def\lsim{\mathrel{\rlap{\lower4pt\hbox{\hskip1pt$\sim$}}
    \raise1pt\hbox{$<$}}}                
\def\gsim{\mathrel{\rlap{\lower4pt\hbox{\hskip1pt$\sim$}}
    \raise1pt\hbox{$>$}}}                
    
\usepackage{graphicx}

\def\sumint{\hbox{$\sum$}\!\!\!\!\!\!\!\int}
\def\square{\vcenter{\vbox{\hrule height.4pt
          \hbox{\vrule width.4pt height8pt
          \kern8pt\vrule width.4pt}\hrule height.4pt}}}

\def\ranglec{\rangle_{\!\!c}}

\newcommand{\beq}{\begin{equation}}
\newcommand{\eeq}{\end{equation}}
\newcommand{\bqa}{\begin{eqnarray}}
\newcommand{\eqa}{\end{eqnarray}}

\title{Three-loop HTL QCD thermodynamics}

\author[a]{Jens O. Andersen,}
\author[a]{Lars E. Leganger,}
\author[b,c]{Michael Strickland}
\author[c,d]{and Nan Su}

\affiliation[a]{Department of Physics, Norwegian University of Science and Technology,\\N-7491 Trondheim, Norway}
\affiliation[b]{Department of Physics, Gettysburg College,\\Gettysburg, PA 17325, 
USA}
\affiliation[c]{Frankfurt Institute for Advanced Studies,\\
Ruth-Moufang-Str. 1,
D-60438 Frankfurt am Main, Germany}
\affiliation[d]{Kavli Institute for Theoretical Physics China,\\ Chinese Academy of
Sciences, Beijing 100190, China}

\emailAdd{andersen@tf.phys.ntnu.no}
\emailAdd{lars.leganger@ntnu.no}
\emailAdd{mstrickl@gettysburg.edu}
\emailAdd{nansu@fias.uni-frankfurt.de}


\abstract{The hard-thermal-loop perturbation theory (HTLpt) framework is used to calculate the thermodynamic functions of a quark-gluon plasma to three-loop order. This is the highest order accessible by finite temperature perturbation theory applied to a non-Abelian gauge theory before the high-temperature infrared catastrophe. All ultraviolet divergences are eliminated by renormalization of the vacuum, the HTL mass parameters, and the strong coupling constant. After choosing a prescription for the mass parameters, the three-loop results for the pressure and trace anomaly are found to be in very good agreement with recent lattice data down to $T \sim 2-3\,T_c$, which are temperatures accessible by current and forthcoming heavy-ion collision experiments.}

\begin{document}
\maketitle

\section{Introduction}

\begin{figure}[t]
\centering
\includegraphics[width=0.6\textwidth]{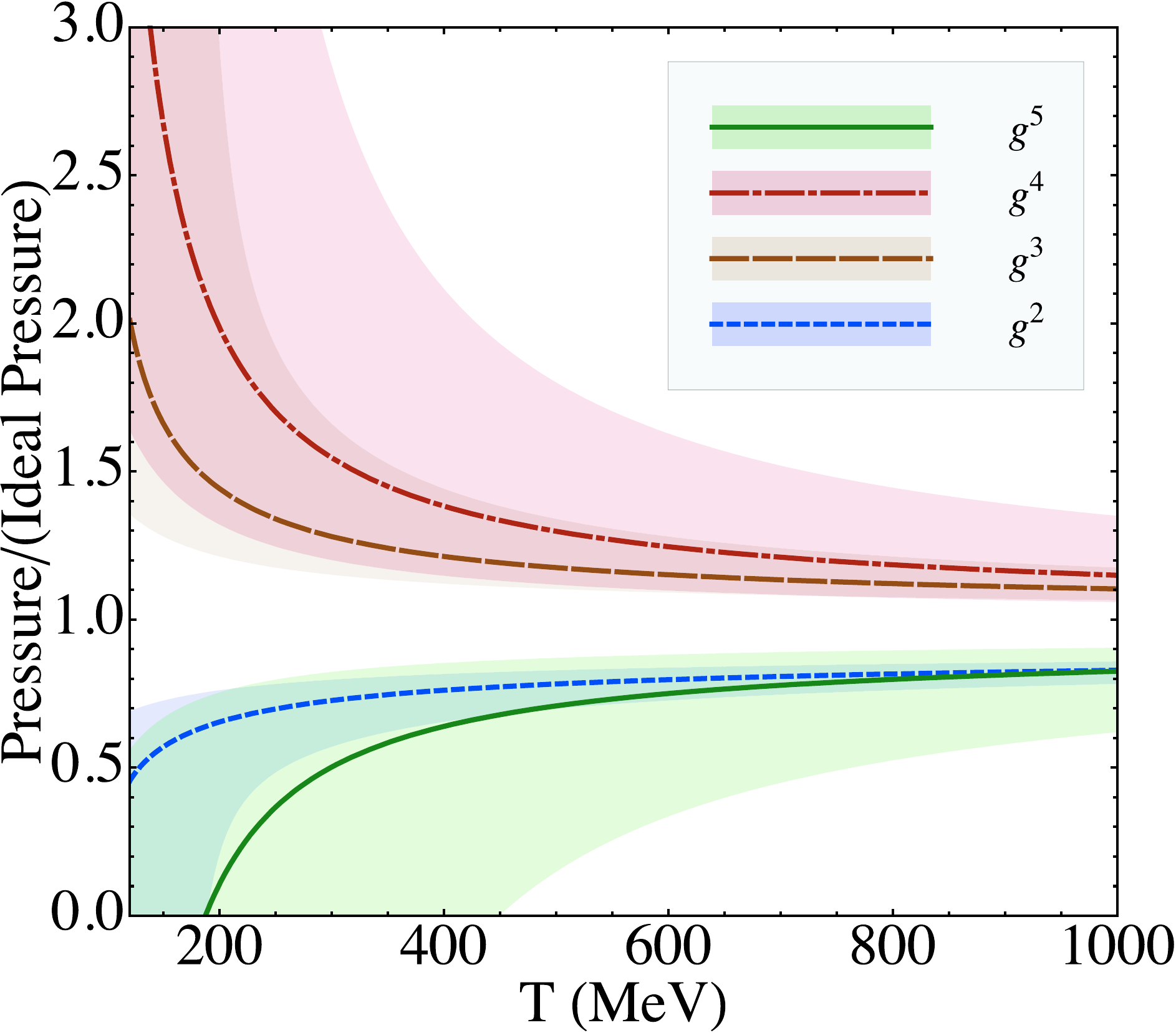}
\caption{Weak-coupling expansion for the scaled QCD pressure with $N_f=3$. Shaded bands show the result of varying the renormalization scale $\mu$ by a factor of 2 around $\mu=2\pi T$.}
\label{fig:pp}
\end{figure}

The last generation of relativistic heavy-ion collision experiments exceeded the energy density necessary for the formation of a quark-gluon plasma, motivating the development of a quantitative framework with which to calculate the properties of this new state of matter. At Brookhaven National Laboratory (RHIC), New York, USA, the initial temperatures were up to twice the critical temperature for deconfinement,\footnote{Note that the deconfinement transition is actually an analytic crossover \cite{Aoki:2006we} and $T_c$ represents a temperature around which the thermodynamic quantities change quickly.}, $T_c \sim170\,$MeV. This translates to a strong coupling constant of $\alpha_s(\mu=2\pi\times170\,{\rm MeV}) \equiv g^2/(4\pi) \sim 0.43$ where $\mu$ is the renormalization scale and it relates to the temperature by $\mu=2\pi T$. The upcoming experiments at CERN (LHC), Geneva, Switzerland, are expected to yield initial temperatures of $4-6\,T_c$, driving the running coupling further down. Initially, the hope was that the asymptotic freedom of QCD would allow calculations using a perturbative expansion in the coupling. Utilizing a weak-coupling expansion in the coupling constant $g$ to calculate the thermodynamic functions of QCD has a long history \cite{Shuryak:1977ut,Kapusta:1979fh,Toimela:1984xy,Arnold:1994ps,Arnold:1994eb,Braaten:1995ju,Braaten:1995jr,Zhai:1995ac}, and the pressure is now known through order $g^6\log g$ for non-Abelian gauge theories \cite{Kajantie:2002wa}. Unfortunately, for all but tiny coupling constants, and thus astronomically high temperatures, these expansions converge very poorly, and show large dependence on the renormalization scale. In figure~\ref{fig:pp}, we show the weak-coupling expansion for the QCD pressure with $N_f=3$ normalized to that of an ideal gas through order $g^5$. The various approximations oscillate wildly and show no signs of convergence in the temperature range shown. The bands are obtained by varying the renormalization scale $\mu$ by a factor of 2 around the value $\mu=2\pi T$ and we use three-loop running for $\alpha_s$~\cite{Amsler:2008zzb} with $\Lambda_{\overline{\rm MS}}(N_f=3)=344\,$MeV~\cite{McNeile:2010ji}. This oscillating behavior is generic for hot field theories, and not specific to QCD. The instability is thought to be caused by plasma effects such as screening of electric fields and Landau damping. This calls for a nonperturbative approach, or a reorganization of the perturbative expansion that takes such effects into account. Furthermore, data from RHIC suggest that the matter created behaves more like a strongly coupled fluid with small viscosity \cite{Arsene:2004fa,Back:2004je,Adams:2005dq,Adcox:2004mh,Gyulassy:2004zy}, inspiring the development of strongly coupled formalisms, perhaps the most successful being those based on the Anti-de-Sitter space/conformal field theory (AdS/CFT) correspondence proposed by Maldacena \cite{Maldacena:1997re}. However, work on the perturbative side has shown that observables like jet quenching~\cite{Qin:2007rn,Qin:2009gw} and elliptic flow~\cite{Xu:2007jv} can also be described by a perturbative setup. Looking forward to the upcoming heavy-ion experiments scheduled to take place at the LHC, it is therefore important to know if, at the higher temperatures generated, one should expect a strongly-coupled (liquid) or weakly-coupled (plasma) description to be more appropriate. The key question is, will the generated matter behave more like a plasma of quasiparticles at these higher temperatures.

Of course, another approach is lattice gauge theory, which is a nonperturbative discrete space-time framework that is the closest one can currently get to a first-principles calculation with realistic parameters. However, the Monte Carlo methods used currently restrict lattice gauge theory to the study of systems with low chemical potential, and reliable calculations of dynamical quantities have proved difficult. A quark-gluon plasma created in a heavy-ion collision will be out of thermal equilibrium and will have nonzero baryon density, making these restrictions quite severe. Still, the part of the phase diagram that is accessible by lattice gauge theory becomes a clean testing ground for the various approaches to analytical QCD calculations. In this paper we will be comparing our new ``reorganized'' perturbative results for the thermodynamic functions to recent lattice data from refs.~\cite{Bazavov:2009zn} and \cite{Borsanyi:2010cj}.

There are several ways of systematically reorganizing the perturbative expansion \cite{Blaizot:2003tw,Kraemmer:2003gd,Andersen:2004fp}. In screened perturbation theory (SPT)~\cite{Karsch:1997gj,Chiku:1998kd,Andersen:2000yj,Andersen:2001ez,Andersen:2008bz} which was inspired by variational perturbation theory (VPT)~\cite{Yukalov:1976pm,Stevenson:1981vj,Duncan:1988hw,Duncan:1992ba,Sisakian:1994nn,Janke:1995zz}, a mass term is added to the free Lagrangian, and then subtracted as an interaction term. Unfortunately, this reorganization cannot be directly applied to gauge theories, since the added local mass term would violate gauge invariance~\cite{Braaten:1991gm,Buchmuller:1994qy,Alexanian:1995rp}. Instead, one can add and subtract a hard-thermal-loop (HTL) effective action~\cite{bp} which modifies the propagators and vertices systematically and self-consistently, so that the reorganization is manifestly gauge invariant \cite{Andersen:1999fw,Andersen:1999sf,Andersen:1999va}. The mass parameters $m_D$ and $m_q$ are introduced, and identified with the Debye screening mass and the thermal quark mass at leading order, respectively, in order to reproduce the high temperature limit of thermal QCD; however, at higher orders it is necessary to introduce a prescription for the subleading contributions to the mass parameters. The resulting HTL perturbation theory (HTLpt) framework can be applied to static as well as dynamic quantities. 

The HTLpt framework has recently been applied to QED \cite{Andersen:2009tw}, where the calculation of the thermodynamic functions is carried out through next-to-next-to-leading-order (NNLO). The thermodynamic functions of QCD have been calculated to next-to-leading-order (NLO) \cite{Andersen:2002ey,Andersen:2003zk}. For pure-glue QCD the thermodynamic functions were recently calculated to NNLO in \cite{Andersen:2009tc,Andersen:2010ct}. Our paper builds upon the results of NNLO QED and pure-glue QCD, and we now include NNLO contributions from quark and quark-gluon interaction diagrams to express the thermodynamic functions of full QCD to NNLO.  The calculation is presented in some detail. For a short letter with just the final results, we refer to ref.~\cite{Andersen:2010wu}. Our results indicate that the lattice data at temperatures $T \sim 2\,T_c$ are consistent with the quasiparticle picture. This is a nontrivial result since, in this temperature regime, the QCD coupling constant is neither infinitesimally weak nor infinitely strong with $g \sim 2$, or equivalently $\alpha_s \sim 0.3$. Therefore, we have a crucial test of the quasiparticle picture in the intermediate coupling regime.

After the introduction, we begin in section~\ref{sec:2} with a summary of HTLpt applied to QCD. Section~\ref{sec:3} contains expressions for the diagrams contributing to the thermodynamic potential up to NNLO. In section~\ref{sec:4} these diagrams are reduced to scalar sum-integrals and expanded in powers of $m_D/T$ and $m_q/T$, keeping all terms that contribute through order $g^5$ if the mass parameters are taken to be of order $g$ at leading order (LO). In section~\ref{sec:5} the calculated diagrams are gathered to obtain the renormalized NNLO thermodynamic potential. The results are then presented in section~\ref{sec:6}, and compared to recent lattice data. Due to difficulties with the normal variational approach, there is special emphasis placed on the determination of the mass parameters $m_D$ and $m_q$. In section~\ref{sec:7} we summarize.

\section{HTL perturbation theory}
\label{sec:2}

\label{HTLpt}

The Lagrangian density for QCD in Minkowski space is
\bqa
{\cal L}_{\rm QCD} =
-{1\over2}{\rm Tr}\left[G_{\mu\nu}G^{\mu\nu}\right]
+i \bar\psi \gamma^\mu D_\mu \psi 
+{\cal L}_{\rm gf}
+{\cal L}_{\rm gh}
+\Delta{\cal L}_{\rm QCD}\;,
\label{L-QCD}
\eqa
where the gluon field strength is $G^{\mu\nu}=\partial^{\mu}A^{\nu}-\partial^{\nu}A^{\mu}-ig[{ A^{\mu},A^{\nu}}]$, the term with the quark fields $\psi$ contains an implicit sum over the $N_f$ quark flavors, and the covariant derivative is $D^{\mu}=\partial^{\mu}-igA^{\mu}$. The ghost term ${\cal L}_{\rm gh}$ depends on the gauge-fixing term ${\cal L}_{\rm gf}$. In this paper we choose the class of covariant gauges where the gauge-fixing term is
\bqa
{\cal L}_{\rm gf} = -{1\over\xi}{\rm Tr}
\left[\left(\partial_{\mu}A^{\mu}\right)^2\right]\;.
\eqa

The perturbative expansion in powers of $g$ generates ultraviolet divergences. The renormalizability of perturbative QCD guarantees that all divergences in physical quantities can be removed by renormalization of the coupling constant $\alpha_s= g^2/(4 \pi)$ and the necessary counterterms are represented by $\Delta{\cal L}_{\rm QCD}$ in the Lagrangian (\ref{L-QCD}). There is no need for wavefunction renormalization, because physical quantities are independent of the normalization of the field. There is also no need for renormalization of the gauge parameter, because physical quantities are independent of the gauge parameter.

Hard-thermal-loop perturbation theory (HTLpt) is a reorganization of the perturbation series for thermal QCD. The Lagrangian density is written as
\bqa
{\cal L}= \left({\cal L}_{\rm QCD}
+ {\cal L}_{\rm HTL} \right) \Big|_{g \to \sqrt{\delta} g}
+ \Delta{\cal L}_{\rm HTL}.
\label{L-HTLQCD}
\eqa
The HTL improvement term is
\bqa
{\cal L}_{\rm HTL}=-{1\over2}(1-\delta)m_D^2 {\rm Tr}
\left(G_{\mu\alpha}\left\langle {y^{\alpha}y^{\beta}\over(y\cdot D)^2}
	\right\rangle_{\!\!y}G^{\mu}_{\;\;\beta}\right)
         +(1-\delta)\,i m_q^2 \bar{\psi}\gamma^\mu 
\left\langle {y_{\mu}\over y\cdot D}
	\right\rangle_{\!\!y}\psi
	\, ,
\label{L-HTL}
\eqa
where $y^{\mu}=(1,\hat{{\bf y}})$ is a light-like four-vector, and $\langle\ldots\rangle_{ y}$ represents the average over the directions of $\hat{{\bf y}}$. The term~(\ref{L-HTL}) has the form of the effective Lagrangian that would be induced by a rotationally invariant ensemble of charged sources with infinitely high momentum and modifies the propagators and vertices self-consistently so that the reorganization is manifestly gauge invariant~\cite{Frenkel:1989br,Taylor:1990ia,Braaten:1991gm,Efraty:1992pd,Jackiw:1993zr,Jackiw:1993pc}. The parameter $m_D$ can be identified with the Debye screening mass, and $m_q$ with the thermal quark mass to account for the screening effects. HTLpt is defined by treating $\delta$ as a formal expansion parameter. By coupling the HTL improvement term~(\ref{L-HTL}) to the QCD Lagrangian~(\ref{L-QCD}), HTLpt systematically shifts the perturbative expansion from being around an ideal gas of massless particles which is the physical picture of the weak-coupling expansion, to being around a gas of massive quasiparticles which are the more appropriate physical degrees of freedom at high temperature.

Physical observables are calculated in HTLpt by expanding them in powers of $\delta$, truncating at some specified order, and then setting $\delta=1$. This defines a reorganization of the perturbation series in which the effects of $m_D^2$ and $m_q^2$ terms in~(\ref{L-HTL}) are included to all orders but then systematically subtracted out at higher orders in perturbation theory by the $\delta m_D^2$ and $\delta m_q^2$ terms in~(\ref{L-HTL}). If we set $\delta=1$, the HTLpt Lagrangian (\ref{L-HTLQCD}) reduces to the QCD Lagrangian (\ref{L-QCD}).

If the expansion in $\delta$ could be calculated to all orders the final result would not depend on $m_D$ and $m_q$ when we set $\delta=1$. However, any truncation of the expansion in $\delta$ produces results that depend on $m_D$ and $m_q$. Some prescription is required to determine $m_D$ and $m_q$ as a function of $T$ and $\alpha_s$. We will discuss several prescriptions in section~\ref{sec:6}.

The HTL perturbation expansion generates ultraviolet divergences. In QCD perturbation theory, renormalizability constrains the ultraviolet divergences to have a form that can be cancelled by the counterterm Lagrangian $\Delta{\cal L}_{\rm QCD}$. We will demonstrate that the renormalization of HTLpt can be implemented by including a counterterm Lagrangian $\Delta{\cal L}_{\rm HTL}$ among the interaction terms in (\ref{L-HTLQCD}). There is no proof that the HTL perturbation expansion is renormalizable, so the general structure of the ultraviolet divergences is not known; however, it was shown in previous papers \cite{Andersen:2002ey,Andersen:2003zk} that it was possible to renormalize the next-to-leading order HTLpt prediction for the pressure of QCD using only a vacuum counterterm, a Debye mass counterterm, and a fermion mass counterterm.  By also including a coupling constant counterterm, this is possible at NNLO as well, as we will show in this paper. The necessary counterterms for the renormalization at NNLO as just discussed read
\bqa
\Delta{\cal E}_0&=&\left({d_A\over128\pi^2\epsilon}
+{\cal O}(\delta\alpha_s)
\right)(1-\delta)^2m_D^4\;,
\label{del1e0}
\\
\Delta m_D^2&=&\left(-{11c_A-4s_F\over12\pi\epsilon}\alpha_s\delta
+{\cal O}(\delta^2\alpha_s^2)
\right)(1-\delta)m_D^2\;,
\label{delmd} \\
\Delta m_q^2&=&\left(-{3\over8\pi\epsilon}{d_A\over c_A}\alpha_s\delta
+{\cal O}(\delta^2\alpha_s^2)
\right)
(1-\delta)m_q^2\;,
\label{delmf}\\
\delta\Delta\alpha_s&=&-{11c_A-4s_F\over12\pi\epsilon}\alpha_s^2\delta^2
+{\cal O}(\delta^3\alpha^3_s)\;,
\label{delalpha}
\eqa
where the vacuum and thermal mass counterterms were derived in ref.~\cite{Andersen:2002ey,Andersen:2003zk}, and the coupling constant counterterm is the standard one-loop running for QCD~\cite{Gross:1973id,Politzer:1973fx}.

\begin{figure}[t]\centering
\includegraphics[trim=102px 240px 102px 66px,clip]{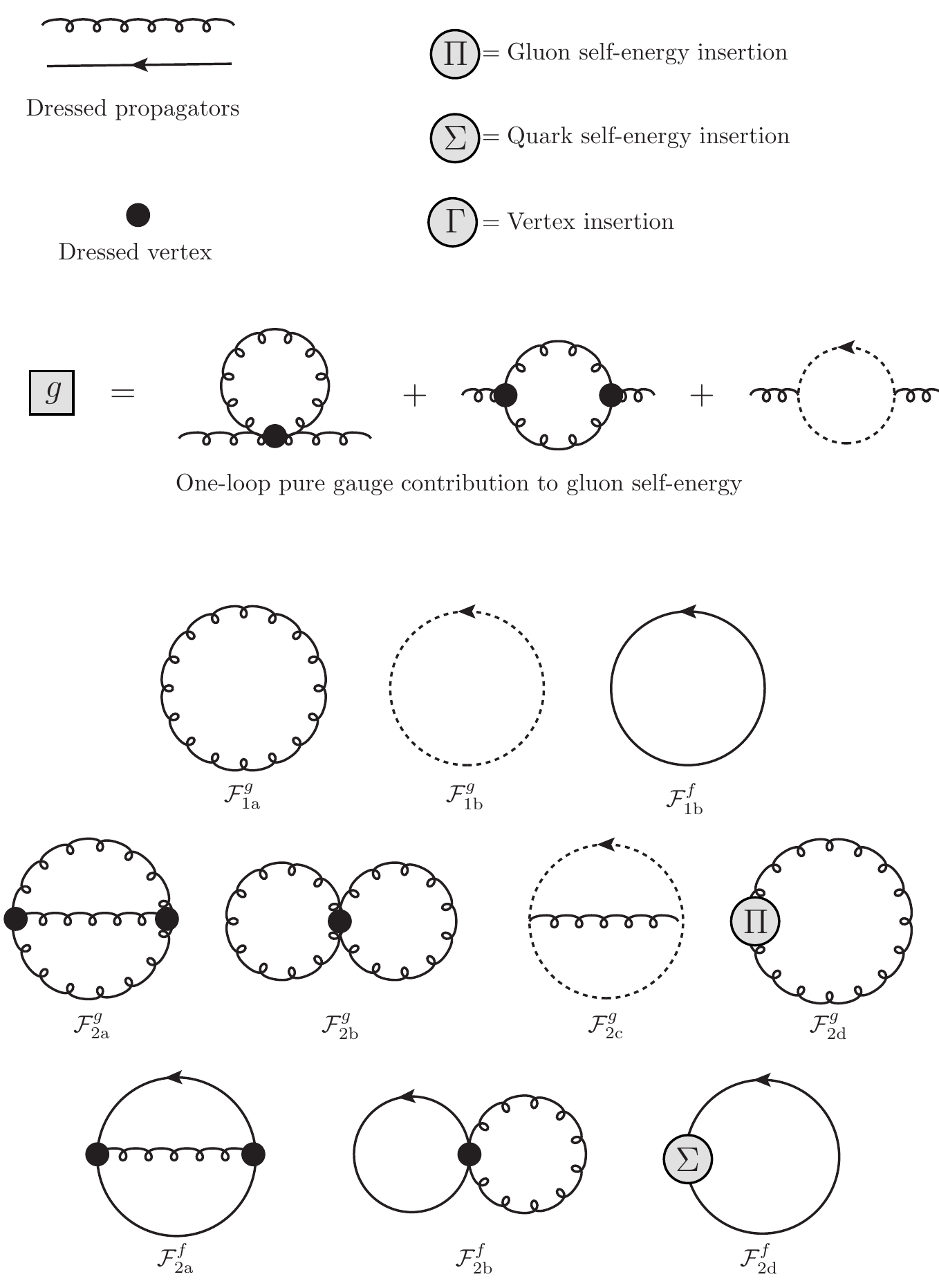}
\caption{QCD diagrams contributing through NLO in HTLpt. The spiral lines are gluon propagators, the dotted lines are ghost propagators, and the solid lines are quark propagators. A circle with a $\Pi$ indicates a one-loop gluon self-energy insertion, a $\Sigma$ indicates a one-loop quark self-energy insertion, and a $\Gamma$ indicates a one-loop vertex insertion. A square with a $g$ is shorthand for the pure gauge diagrams contributing to the one-loop gluon self-energy. All gluon and quark propagators and vertices shown are HTL-resummed propagators and vertices. The logic behind the diagram notation is as follows: diagrams consisting only of gauge propagators have $g$ superscripts. Diagrams containing fermion propagators have $f$ superscripts. The subscript indices are identical to those used in \cite{Andersen:2010ct} (pure gauge QCD) and \cite{Andersen:2009tw} (QED). We do not display the symmetry factors in the diagrams.}
\label{fig:LO&NLO}
\end{figure}

\pagebreak
\clearpage

\begin{figure}[b]\centering
\includegraphics[trim=104px 169px 95px 23px,clip]{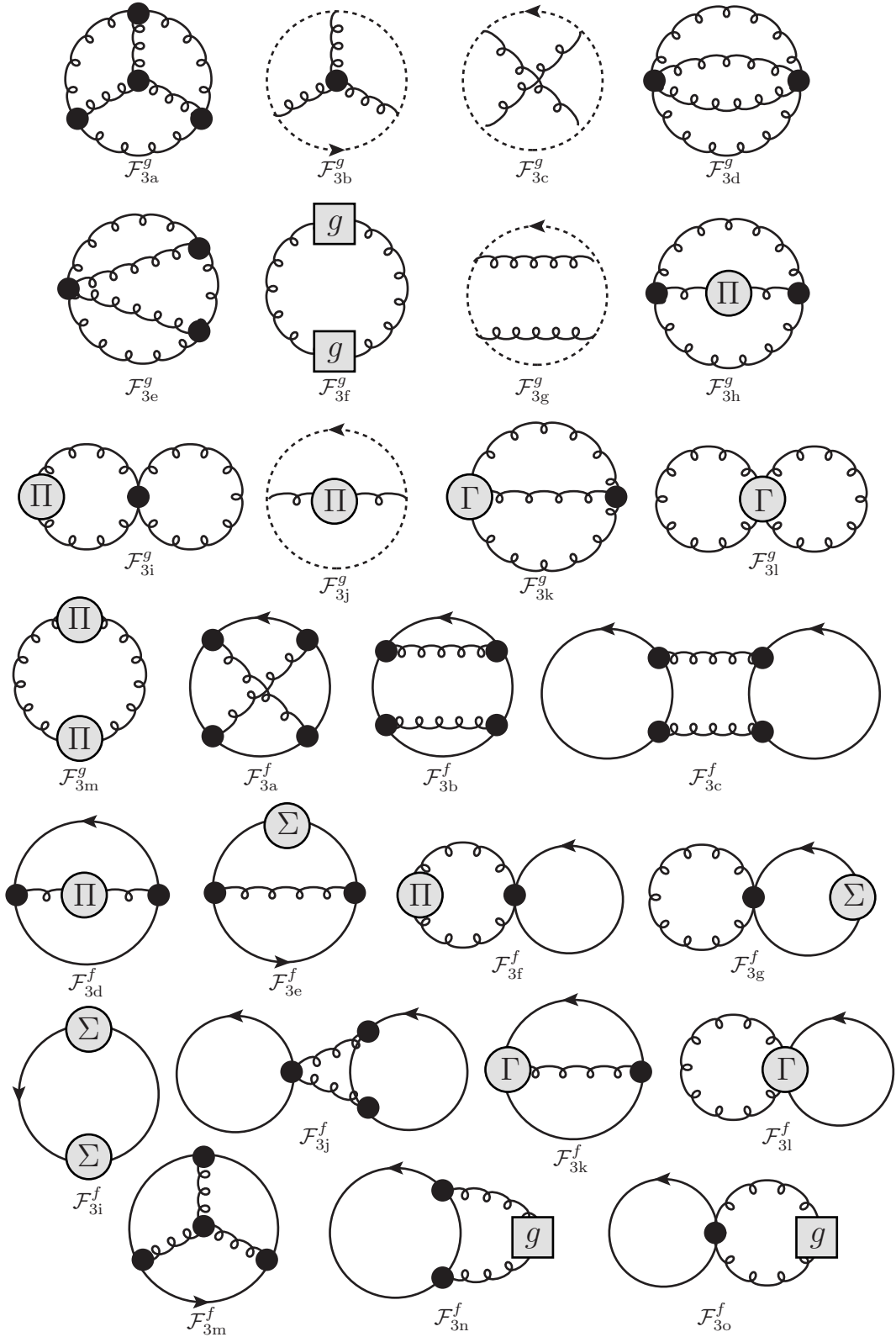}
\caption{QCD diagrams contributing to NNLO in HTLpt.}
\label{fig:NNLO}
\end{figure}
\pagebreak
\clearpage

\section{Diagrams for the thermodynamic potential}
\label{sec:3}

Although HTLpt is defined in Minkowski space, it is much more convenient to carry out the necessary calculations in Euclidean space since in the paper we focus on static quantities. In the imaginary-time formalism, Minkoswski energies have discrete imaginary values $p_0 = i 2 n \pi T$ for bosons and $p_0 = i (2n+1) 2 \pi T$ for fermions, and integrals over Minkowski space are replaced by sum-integrals over Euclidean vectors $(2 n \pi T, {\bf p})$ or $(2 (n+1) \pi T, {\bf p})$, respectively, with $n$ an integer. We will use the notation $P=(P_0,{\bf p})$ for Euclidean momenta. The magnitude of the spatial momentum will be denoted $p = |{\bf p}|$. The inner product of two Euclidean vectors is $P \cdot Q = P_0 Q_0 + {\bf p} \cdot {\bf q}$. The vector that specifies the thermal rest frame remains $n = (1,{\bf 0})$. Here we use dimensional regularization and the remaining three-dimensional integral is generalized to $d = 3-2 \epsilon$ spatial dimensions. We define the dimensionally regularized sum-integral by
\bqa
  \hbox{$\sum$}\!\!\!\!\!\!\int_{P} &\equiv& 
  \left(\frac{e^{ \gamma_E}\mu^2}{4\pi}\right)^\epsilon\;
  T\sum_{P_0=2n\pi T}\:\int {d^{3-2\epsilon}p \over (2 \pi)^{3-2\epsilon}}
\hspace{2.7cm}\mbox{bosons} \;,
\label{sumint-b}\\ 
  \hbox{$\sum$}\!\!\!\!\!\!\int_{\{P\}} &\equiv&
  \left(\frac{e^{ \gamma_E}\mu^2}{4\pi}\right)^\epsilon\;
  T\sum_{P_0=(2n+1)\pi T}\:\int {d^{3-2\epsilon}p \over (2 \pi)^{3-2\epsilon}}
\hspace{2cm}\mbox{fermions} \;,
\label{sumint-def}
\eqa
where $\mu$ is an arbitrary momentum scale. The factor $(e^{ \gamma_E}/4\pi)^\epsilon$ is introduced so that, after minimal subtraction of the poles in $\epsilon$ due to ultraviolet divergences, $\mu$ coincides with the renormalization scale of the $\overline{\rm MS}$ renormalization scheme.

The Feynman diagrams discussed in this section are gathered in figures~\ref{fig:LO&NLO} (notation key, LO, NLO) and \ref{fig:NNLO} (NNLO). Using the same notation for the group theory factors as Arnold and Zhai \cite{Arnold:1994eb}, for QCD with $N_c$ colors and $N_f$ flavors of quarks, we have
\beq
d_A 	= N_c^2-1\;,			\qquad
d_F 	= N_c N_f\;, 			\qquad
c_A		= N_c\;,				\qquad
s_F		= {N_f\over2}\;,		\qquad
s_{2F}	= {N_c^2-1\over4N_c}N_f\;.
\label{casimirs}
\eeq
The earlier work on NNLO pure-glue QCD \cite{Andersen:2010ct} and NNLO QED \cite{Andersen:2009tw} essentially treat special cases where the group symmetry factors are
\beq
d_A 	= N_c^2-1\;,		\qquad
d_F 	= 0\;, 			\qquad
c_A		= N_c\;,			\qquad
s_F		= 0\;,		\qquad
s_{2F}	= 0\;,
 \;\;\;\textrm{(pure-glue QCD)}
\eeq
or
\beq
d_A 	= 1\;,			\qquad
d_F 	= N_f\;, 			\qquad
c_A		= 0\;,			\qquad
s_F		= N_f\;,		\qquad
s_{2F}	= N_f\;,
 \;\;\;\textrm{(QED)}
\eeq
and as such the analytical results from those papers will be highly useful in the present calculation as well.

The thermodynamic potential at LO in HTLpt for QCD reads
\bqa
\Omega_{\rm LO}= 
d_A\left[{\cal F}_{\rm 1a}^g+{\cal F}_{\rm 1b}^g\right]
+ d_F{\cal F}_{\rm 1b}^f
+\Delta_0{\cal E}_0\;.
\label{OmegaLO}
\eqa
Here, ${\cal F}^g_{\rm 1a}$ is the contribution from the gluons and ${\cal F}^g_{\rm 1b}$ the contribution from the ghost, and $\Delta_0{\cal E}_0$ is the leading-order vacuum counterterm. The expression of eq.~(\ref{OmegaLO}) is presented in ref.~\cite{Andersen:2003zk}, and the details can be found in section 3 therein.

The thermodynamic potential at NLO reads
\bqa\nonumber
\Omega_{\rm NLO}&=&\Omega_{\rm LO}
 + d_Ac_A\left[{\cal F}_{\rm 2a}^g+{\cal F}_{\rm 2b}^g+{\cal F}_{\rm 2c}^g
\right]
+d_A{\cal F}_{\rm 2d}^g
+d_As_F
\left[
{\cal F}^f_{\rm 2a}+{\cal F}_{\rm 2b}^f
\right]
+d_F{\cal F}_{\rm 2d}^f
\\&&
+\Delta_1{\cal E}_0
+\Delta_1 m_D^2{\partial\over\partial m_D^2}
\Omega_{\rm LO}
+\Delta_1 m_q^2{\partial\over\partial m_q^2}\Omega_{\rm LO}
\;,
\label{OmegaNLO}
\eqa
where $\Delta_1{\cal E}_0$, $\Delta_1m_D^2$, and 
$\Delta_1m_q^2$ 
are the terms of order
$\delta$ in the vacuum energy density and mass counterterms.
Again the detailed expression of eq.~(\ref{OmegaNLO}) is presented in section 3 of ref.~\cite{Andersen:2003zk}.

The NNLO HTLpt thermodynamic potential for QCD can be written as
\bqa\nonumber
\Omega_{\rm NNLO} &=&
\Omega_{\rm NLO}
+d_Ac_A^2\left[
{\cal F}^g_{\rm 3a}+{\cal F}^g_{\rm 3b}+{\cal F}^g_{\rm 3c}
+{\cal F}^g_{\rm 3d}+{\cal F}^g_{\rm 3e}
+{\cal F}^g_{\rm 3f}+{\cal F}^g_{\rm 3g}
\right]
\\ \nonumber &&
+d_Ac_A\left[
{\cal F}^g_{\rm 3h}+{\cal F}^g_{\rm 3i}
+{\cal F}^g_{\rm 3j}+{\cal F}^g_{\rm 3k}+{\cal F}^g_{\rm 3l}
\right]
+d_As_{2F}\left[{\cal F}^f_{\rm 3a}
+{\cal F}^f_{\rm 3b}\right]
\\ &&\nonumber
+d_As_F^2\left[{\cal F}^f_{\rm 3c}
+{\cal F}_{\rm 3j}^f\right]
+d_Ac_As_F\left[-{1\over2}{\cal F}^f_{\rm 3a}
+{\cal F}_{\rm 3m}^f+{\cal F}_{\rm 3n}^f+{\cal F}_{\rm 3o}^f
\right]
\\ &&\nonumber
+d_As_F\left[
{\cal F}^f_{\rm 3d}+{\cal F}^f_{\rm 3e}
+{\cal F}^f_{\rm 3f}+{\cal F}^f_{\rm 3g}
+{\cal F}^f_{\rm 3k}+{\cal F}^f_{\rm 3l}
\right]
\\ &&\nonumber
+d_A{\cal F}^g_{\rm 3m}+d_F{\cal F}^f_{\rm 3i}
+\Delta_2{\cal E}_0
+\Delta_2 m_D^2{\partial\over\partial m_D^2}
\Omega_{\rm LO}
+\Delta_2 m_q^2{\partial\over\partial m_q^2}\Omega_{\rm LO}
\\ &&\nonumber
+\Delta_1 m_D^2{\partial\over\partial m_D^2}
\Omega_{\rm NLO}
+\Delta_1 m_q^2{\partial\over\partial m_q^2}\Omega_{\rm NLO}
+{1\over2}\left[{\partial^2\over(\partial m_D^2)^2}
\Omega_{\rm LO}\right]\left(\Delta_1m_D^2\right)^2
\\ &&
+{1\over2}\left[{\partial^2\over(\partial m_q^2)^2}
\Omega_{\rm LO}\right]\left(\Delta_1m_q^2\right)^2
+d_A\left[{c_A{\cal F}^g_{\rm 2a+2b+2c}+s_F{\cal F}_{\rm 2a+2b}^f
\over\alpha_s}\right]\Delta_2\alpha_s
\;,\nonumber\\
\label{OmegaNNLO}
\eqa
where $\Delta_2{\cal E}_0$, $\Delta_2m_D^2$, $\Delta_2m_q^2$ and $\Delta_2\alpha_s$ are the order $\delta^2$ counterterms in the vacuum energy density, masses and coupling constant that can be read out from eqs.~(\ref{del1e0}),~(\ref{delmd}),~(\ref{delmf}), and~(\ref{delalpha}).

The expressions for the bosonic diagrams ${\cal F}_{3a}^g$--${\cal F}_{3m}^g$ are presented in section 3 of ref.~\cite{Andersen:2010ct}, and the diagrams with fermions ${\cal F}_{3a}^f$--${\cal F}_{3i}^f$ are expressed in section 3 of ref.~\cite{Andersen:2009tw}. The only NNLO diagrams specific to QCD, i.e. the non-Abelian diagrams involving quarks, are given by
\bqa\nonumber
{\cal F}_{\rm 3m}^f
&=&{1\over3}g^4
\sumint_{\{PQR\}}{\rm Tr}\left[
\Gamma^{\alpha}(R-P,R,P)S(P)\Gamma^{\beta}(P-Q,P,Q)S(Q)
\Gamma^{\gamma}(Q-R,Q,R)S(R)\right]
\\ &&\times
\Gamma^{\mu\nu\delta}(P-R,Q-P,R-Q)
\Delta^{\alpha\mu}(P-R)\Delta^{\beta\nu}(Q-P)\Delta^{\gamma\delta}(R-Q)\;,
\\
{\cal F}^f_{\rm 3n}&=&
{-}\sumint_{P}\bar{\Pi}_g^{\mu\nu}(P)\Delta^{\nu\alpha}(P)
\bar{\Pi}_f^{\alpha\beta}(P)\Delta^{\beta\mu}(P)\;,
\\
{\cal F}^f_{\rm 3o}&=&
-{1\over2}
g^2\sumint_{P\{Q\}}
{\rm Tr}\left[
\Gamma^{\alpha\beta}(P,-P,Q,Q)
S(Q)\right]\Delta^{\alpha\mu}(P)\Delta^{\beta\nu}(P)
\bar{\Pi}_g^{\mu\nu}(P)\;,
\eqa
where
\bqa
\nonumber
\bar{\Pi}^{\mu\nu}_g(P)
&=&{1\over2}g^2\sumint_Q
\Gamma^{\mu\nu,\alpha\beta}(P,-P,Q,-Q)\Delta^{\alpha\beta}(Q)
\\ && \nonumber
+{1\over2}g^2\sumint_Q\Gamma^{\mu\alpha\beta}(P,Q,-P-Q)\Delta^{\alpha\beta}(Q)
\Gamma^{\nu\gamma\delta}(P,Q,-P-Q)\Delta^{\gamma\delta}(-P-Q)
\\&&
+g^2\sumint_Q{Q^{\mu}(P+Q)^{\nu}\over Q^2(P+Q)^2}\;,
\\ 
\bar{\Pi}^{\mu\nu}_f(P)
&=&
{-}g^2\sumint_{\{Q\}}{\rm Tr}\left[\Gamma^{\mu}(P,Q,Q-P)
S(Q)\Gamma^{\nu}(P,Q,Q-P)S(Q-P)\right]
\;.
\eqa
Thus $\bar{\Pi}^{\mu\nu}(P)$ is the one-loop gluon self-energy with
HTL-resummed propagators and vertices:
\bqa
\bar{\Pi}^{\mu\nu}(P) = 
c_A\bar{\Pi}^{\mu\nu}_g(P)+s_F\bar{\Pi}^{\mu\nu}_f(P)\;.
\eqa

\section{Expansion in the mass parameters}
\label{sec:4}
In refs.~\cite{Andersen:2002ey,Andersen:2003zk}, the NLO HTLpt pressure was reduced to scalar sum-integrals. It was clear that evaluating these scalar sum-integrals exactly was intractable and the sum-integrals were calculated approximately by expanding them in powers of $m_D/T$ and $m_q/T$ following the method developed in ref.~\cite{Andersen:2001ez}. We will adopt the same strategy in this paper and carry out the expansion to high enough order to include all terms through order $g^5$ if $m_D$ and $m_q$ are taken to be of order $g$.

The pressure can be divided into contributions from hard and soft momenta, which are the momenta proportional to the scales $T$ and $gT$ respectively. In the one-loop diagrams, the contributions are either hard $(h)$ or soft $(s)$, while at the two-loop level, there are hard-hard $(hh)$, hard-soft $(hs)$, and soft-soft $(ss)$ contributions. At three loops there are hard-hard-hard $(hhh)$, hard-hard-soft $(hhs)$, hard-soft-soft $(hss)$, and soft-soft-soft $(sss)$ contributions. As mentioned above, the $(h)$, $(s)$, $(hh)$, $(hs)$ and $(ss)$ contributions have been calculated in refs.~\cite{Andersen:2002ey,Andersen:2003zk}, while most of the three-loop contributions can be read out from the earlier NNLO HTLpt work on pure-glue QCD~\cite{Andersen:2010ct} and QED~\cite{Andersen:2009tw}, thus to keep the discussion compact, we only list the contribution that is specific to QCD here, i.e the term proportional to $c_As_F$ in eq.~(\ref{OmegaNNLO}).

If all momenta in the three loops are hard, we can obtain the $m_D/T$ and $m_q/T$ expansion by expanding in powers of $m_D^2$ and $m_q^2$, and in this case to obtain the expansion through order $g^5$, we can simply use bare propagators and vertices. The contributions from the three-loop diagrams were first calculated by Arnold and Zhai in ref.~\cite{Arnold:1994ps}, and later by Braaten and Nieto in ref.~\cite{Braaten:1995jr} from which the $(hhh)$ contribution to the $c_A s_F$ term reads
\bqa
\nonumber
&& - {1\over2}{\cal F}^{f(hhh)}_{\rm 3a} + {\cal F}^{f(hhh)}_{\rm 3m+3n}
\\ \nonumber
&=&g^4(d-1)\Bigg\{2{ (d-5)}\sumint_{PQ\{R\}}
{1\over P^4Q^2R^2}
+{1\over2}(d-3)\sumint_{\{PQR\}}
{1\over P^2Q^2(P-Q)^2(Q-R)^2}
\\&& \nonumber
\hspace{2cm}
-{1\over4}(d-5)\sumint_{\{PQ\}R}
{1\over P^2Q^2(P-Q)^2(Q-R)^2}
\\&& \nonumber
\hspace{2cm}
+2\sumint_{\{PQR\}}{R^4\over P^2Q^2(P-Q)^4(Q-R)^2(R-P)^2}
\Bigg\}
\\
&=&
-{25\alpha_s^2T^4\over864}
\left[
        {1\over\epsilon}
        { - {369\over250}}
        -{282\over125}\log2
        +{48\over25}\gamma_E
        +{104\over25}{\zeta'(-1)\over\zeta(-1)}
        -{2\over25}{\zeta'(-3)\over\zeta(-3)}
\right]
\left({\mu\over4\pi T}\right)^{6\epsilon}
\;.
\eqa
Note that there is no $(hhh)$ contribution from the HTL-specific diagram ${\cal F}^{f}_{\rm 3o}$~\cite{Andersen:2009tw}.

In the $(hhs)$ region, the momentum $P$ is soft,\footnote{The soft contribution arises from the $P_0=0$ term in the sum-integral. See appendix A of ref.~\cite{Braaten:2001vr} for details.} while the momenta $Q$ and $R$ are always hard. The function that multiplies the static propagator $\Delta_T(0,{\bf p})$, $\Delta_L(0,{\bf p})$ or $\Delta_X(0,{\bf p})$ can be expanded in powers of the momentum ${\bf p}$. In the case of $\Delta_T(0,{\bf p})$, the resulting integrals over ${\bf p}$ have no scale and thus vanish in dimensional regularization. The integration measure $\int_{\bf p}$ scales like $m_D^3$, the static propagators $\Delta_L(0,{\bf p})$ and $\Delta_X(0,{\bf p})$ scale like $1/m_D^2$, and every power of $p$ in the numerator scales like $m_D$.\footnote{We refer the reader to e.g. ref.~\cite{Andersen:2002ey} for HTLpt Feynman rules, notations and conventions.} The $(hhs)$ contribution to the $c_A s_F$ term reads
\bqa\nonumber
&& - {1\over2}{\cal F}^{f(hhs)}_{\rm 3a} + {\cal F}^{f(hhs)}_{\rm 3m+3n+3o}
\\ \nonumber
&=&
g^4T(d-1)\Bigg\{2(d-1)^2\int_p{1 \over (p^2+m_D^2)^2}\sumint_{Q\{R\}}{1\over Q^2R^2}
\\ \nonumber &&
\hspace{2.2cm}
-{1\over3}(d^2-11d+46)
\int_p{p^2\over p^2+m_D^2}
\sumint_{Q\{R\}}{1\over Q^4R^2}
\\ &&\nonumber
\hspace{2.2cm}
-{1\over3}(d-1)^2\int_p{p^2 \over (p^2+m_D^2)^2}
\sumint_{Q\{R\}}{1\over Q^2R^4}
\\ &&\nonumber
\hspace{2.2cm}
+4m_q^2(d-1)\int_{p}{1\over(p^2+m_D^2)^2}
\\&&\nonumber
\hspace{2.5cm}
\times
\sumint_{Q}{1\over Q^2}
\sumint_{\{R\}}\left[
{3\over R^4}-{4r^2\over R^6}
-{4\over R^4}{\cal T}_R
-{2\over R^2}
\bigg\langle{1\over(R\!\cdot\!Y)^2} \bigg\rangle_{\!\!\bf \hat y}\right] \Bigg\}\\
&=& \nonumber
-{\pi\alpha_s^2T^5\over9m_D}
-{ }{7\alpha_s^2m_DT^3\over48\pi}
\left[
{1\over\epsilon}+{88\over21}+2\gamma_E-{38\over7}\log2
+2{\zeta'(-1)\over\zeta(-1)}
\right]
\left({\mu\over4\pi T}\right)^{4\epsilon}
\left({\mu\over2m_D}\right)^{2\epsilon} 
\\ &&
-{\alpha_s^2\over3\pi m_D}m_q^2T^3
\;,
\eqa
with the function ${\cal T}_P$ defined by
\bqa
{\cal T}_P =
        \left\langle {P_0^2 \over P_0^2 + p^2c^2} \right\ranglec \;,
\label{TP-int}
\eqa
where the angular brackets represent an average over $c$
\begin{equation}
\left\langle f(c) \right\rangle_{\!c} \equiv {\Gamma({3\over2}-\epsilon)
        \over \Gamma({3\over2}) \Gamma(1-\epsilon)}
        \int_0^1 dc \, (1-c^2)^{-\epsilon} f(c) \;.
\label{c-average}
\end{equation}

For all of the diagrams that are infrared safe, the $(hss)$ contribution is of order $g^4m_D^2$, i.e. $g^6$, and can be ignored. The infrared divergent diagrams, i.e. ${\cal F}^{f}_{\rm 3n}$, contribute to the $c_As_F$ term as follow
\bqa\nonumber
{\cal F}^{f(hss)}_{\rm 3n}&=&
-g^4T^2(d-1)\sumint_{\{R\}}{1\over R^2}
\int_{pq}
\left[
{4\over p^2(q^2+m^2_D)(r^2+m_D^2)}
-{2(p^2+4m_D^2)\over p^2(q^2+m^2_D)^2(r^2+m_D^2)}
\right] \\
&=&
{\alpha_s^2T^4\over12}
\left({1\over\epsilon}+2-2\log2+2{\zeta^{\prime}(-1)\over\zeta(-1)}\right)
\left({\mu\over2m_D}\right)^{4\epsilon}\left({\mu\over4\pi T}\right)^{2\epsilon}
\;.
\eqa

\section{The thermodynamic potential}
\label{sec:5}

In this section we present the final renormalized QCD thermodynamic potential. The LO and NLO results were derived in refs.~\cite{Andersen:2002ey,Andersen:2003zk}, thus to keep the presentation compact, we only list the NNLO one, which corresponds to order $\delta^2$ in HTLpt.

Using the results listed in section~\ref{sec:4}, the renormalization contributions at order $\delta^2$ read
\bqa\nonumber
\Delta\Omega_2&=&\Delta_2{\cal E}_0
+\Delta_2m_D^2{\partial\over\partial m_D^2}\Omega_{\rm LO}
+\Delta_2m_q^2{\partial\over\partial m_q^2}\Omega_{\rm LO}
+\Delta_1m_D^2{\partial\over\partial m_D^2}\Omega_{\rm NLO}
+\Delta_1m_q^2{\partial\over\partial m_q^2}\Omega_{\rm NLO}
\\ && \nonumber
+{1\over2}\left({\partial^2\over(\partial m_D^2)^2}
\Omega_{\rm LO}\right)\left(\Delta_1m_D^2\right)^2
+{1\over2}\left({\partial^2\over(\partial m_q^2)^2}
\Omega_{\rm LO}\right)\left(\Delta_1m_q^2\right)^2
\nonumber\\&& \nonumber
+d_A\left[{c_A{\cal F}^g_{\rm 2a+2b+2c}
+{ s}_F{\cal F}_{\rm 2a+2b}^f\over\alpha_s}\right]\Delta_2\alpha_s
\\ \nonumber
&=&
{\cal F}_{\rm ideal}
\Bigg\{ 
{s_F\alpha_s\over\pi}\left[
{5\over2}
\left(
{1\over\epsilon}+2\log{\hat\mu\over 2} +2+ 2 {\zeta'(-1)\over\zeta(-1)}\right)
\hat m_D^2
\right.\\ &&\nonumber\left.
\hspace{1.2cm}
-{45\over2}
\left(
{1\over\epsilon}
+2\log{\hat\mu\over 2} -2\log{\hat{m}_D}+{4\over3}
\right)\hat{m}_D^3
\right.\\ &&\nonumber\left.
\hspace{1.2cm}
-{45\over2}
\left(
{1\over\epsilon}+2+
2\log{\hat\mu\over 2} -2\log2+ 2 {\zeta'(-1)\over\zeta(-1)}
\right)\hat{m}_q^2\right]
\\ &&\nonumber
\hspace{1.2cm}
+\left({c_A\alpha_s\over3\pi}\right)\left({s_F\alpha_s\over\pi}\right)
\left[{235\over32}\left({1\over\epsilon}+4\log{\hat{\mu}\over2}
+4{\zeta^{\prime}(-1)\over\zeta(-1)}
+{149\over47}-{132\over47}\log2
\right)
\right.\\ &&\nonumber\left.
\hspace{1.2cm}
-{315\over8}\left(
{1\over\epsilon}
+4\log{\hat{\mu}\over2}-2\log{\hat{m}_D}
+{61\over21}
-{22\over7}\log2
+2{\zeta^{\prime}(-1)\over\zeta(-1)}
\right)\hat{m}_D
\right]
\\ &&\nonumber
\hspace{1.2cm}
+\left({s_F\alpha_s\over\pi}\right)^2\left[
-{25\over24}\left({1\over\epsilon}+4\log{\hat{\mu}\over2}
+3-{12\over5}\log2+4{\zeta^{\prime}(-1)\over\zeta(-1)}
\right)
\right.\\&&\left.
\hspace{1.2cm}
+{15\over2}\left({1\over\epsilon}
+4\log{\hat{\mu}\over2}-2\log{\hat{m}_D}
+{7\over3}-2\log2+2{\zeta^{\prime}(-1)\over\zeta(-1)}
\right)\hat{m}_D
\right]
\Bigg\} 
+\Delta\Omega_2^{\rm YM}
\;,
\nonumber \\
\label{OmegaVMct2}
\eqa
where ${\cal F}_{\rm ideal}$ is the pressure of a gas of $d_A$ massless spin-one bosons and $\hat m_D$, $\hat m_q$ and $\hat \mu$ are dimensionless variables:
\bqa
{\cal F}_{\rm ideal}&=&d_A\left(-{\pi^2\over45}T^4\right)\;,\\
\hat m_D &=& {m_D \over 2 \pi T}  \;,
\\
\hat m_q &=& {m_q \over 2 \pi T}  \;,
\\
\hat \mu &=& {\mu \over 2 \pi T}  \;. 
\eqa
In order to keep the expression (\ref{OmegaVMct2}) compact, the pure-glue contributions have been represented by $\Delta\Omega_2^{\rm YM}$ which is given by eq.~(4.10) of ref.~\cite{Andersen:2010ct}.

Adding the NNLO counterterms (\ref{OmegaVMct2}) to the contributions from the various NNLO diagrams we obtain the renormalized NNLO thermodynamic potential. We note that at NNLO all numerically determined subleading coefficients in $\epsilon$ drop out and due to this the final result is completely analytic. The resulting NNLO thermodynamic potential is
\begin{eqnarray}\nonumber
\Omega_{\rm NNLO}&=&
{\cal F}_{\rm ideal}
\bigg\{
{7\over4}{d_F\over d_A}
+{s_F\alpha_s\over\pi}\bigg[-{25\over8}+{15\over2}\hat{m}_D
+15\left(
\log{\hat\mu \over 2}-{1\over2}+\gamma_E+2\log2\right)\hat m_D^3
\\&& \nonumber
\hspace{1.2cm}
-90\hat{m}^2_{ q}\hat{m}_D\bigg]
+\left({c_A\alpha_s\over3\pi}\right)\left({s_F\alpha_s\over\pi}\right)
\bigg[
{15\over2}{1\over\hat{m}_D}
-{235\over16}\bigg(\log{\hat{\mu}\over2}
-{144\over47}\log{\hat{m}_D}
\\ && \nonumber
\hspace{1.2cm}
-{24\over47}\gamma_E
+{319\over940}+{111\over235}\log2
-{74\over47}{\zeta^{\prime}(-1)\over\zeta(-1)}
+{1\over47}{\zeta^{\prime}(-3)\over\zeta(-3)}
\bigg)
\\ &&\nonumber
\hspace{1.2cm}
+{315\over4}\bigg(\log{\hat{\mu}\over2}
-{8\over7}\log2+\gamma_E+{9\over14}
\bigg)\hat{m}_D
+90{\hat{m}_{ q}^2\over\hat{m}_D}
\bigg]
+\left({s_F\alpha_s\over\pi}\right)^2
\bigg[{5\over4}{1\over\hat{m}_D}
\\ &&\nonumber
\hspace{1.2cm}
+{25\over12}\bigg(
\log{\hat{\mu}\over2}
+{1\over20}+{3\over5}\gamma_E-{66\over25}\log2
+{4\over5}{\zeta^{\prime}(-1)\over\zeta(-1)}
-{2\over5}{\zeta^{\prime}(-3)\over\zeta(-3)}\bigg)
\\ &&
\hspace{1.2cm}
-15\left(\log{\hat{\mu}\over2}
-{1\over2}+\gamma_E+2\log2
\right)\hat{m}_D
+30{\hat{m}_{ q}^2\over\hat{m}_D}
\bigg]
\nonumber\\&&
\hspace{1.2cm}
+s_{2F}\left({\alpha_s\over\pi}\right)^2\left[{15\over64}(35-32\log2)
-{45\over2}\hat{m}_D\right]
\bigg\}
+\Omega_{\rm NNLO}^{\rm YM}
\;,
\label{Omega-NNLO}
\end{eqnarray}
where $\Omega_{\rm NNLO}^{\rm YM}$ is the NNLO pure-glue thermodynamic potential given by eq.~(4.11) in ref.~\cite{Andersen:2010ct} We note that the coupling constant counterterm listed in eq.~(\ref{delalpha}) coincides with the known one-loop running of QCD
\beq
\mu \frac{d g^2}{d \mu} = -\frac{(11c_A-4s_F)g^4}{24\pi^2} \;.
\label{runningcoupling}
\eeq
In the next section we will present results as a function of $g$ evaluated at the renormalization scale $2 \pi T$.

\section{Thermodynamic functions}
\label{sec:6}

The mass parameters $m_D$ and $m_q$ in HTLpt are completely arbitrary. To complete a 
calculation, it is necessary to specify $m_D$ and $m_q$ as
functions of $g$ and $T$.
As HTLpt is inspired by variational perturbation theory, the natural prescription for specifying the mass parameters is to use the variational solutions resulting from the gap equations
\bqa
{\partial \ \ \over \partial m_D}\Omega(T,\alpha_s,m_D,m_q,\mu,\delta=1) 
&=& 0 \;,
\label{gapmd}\\
{\partial \ \ \over \partial m_q}\Omega(T,\alpha_s,m_D,m_q,\mu,\delta=1) &=& 0 
\;.
\label{gapmq}
\eqa
The LO gap equations are not well defined due to the lack of the coupling constant $g$ in the LO thermodynamic potential, and in previous works on LO HTLpt one therefore chose the LO weak-coupling perturbative values of the Debye and thermal quark masses as physical estimates of the HTLpt mass parameters \cite{Andersen:1999fw,Andersen:1999sf,Andersen:1999va}. At NLO the variational prescription gives nontrivial solutions for the mass parameters, and the NLO results presented in \cite{Andersen:2002ey,Andersen:2003zk} use this prescription for both $m_D$ and $m_q$. For our NNLO results the variational prescription yields a complex Debye mass and $m_q=0$. The problem of a complex mass parameter was encountered at NNLO in SPT massless $\phi^4$ theory \cite{Andersen:2000yj}, in HTLpt QED \cite{Andersen:2009tw}, and in HTLpt pure-glue QCD \cite{Andersen:2009tc,Andersen:2010ct}, so this seems to be a general issue with SPT/HTLpt at NNLO, which is not specific to QCD. Since the weak-coupling perturbative Debye mass receives contributions from the nonperturbative magnetic scale beyond LO, for the NNLO HTLpt pure-glue QCD results the Braaten and Nieto's (BN) mass parameter of three-dimensional electrostatic QCD (EQCD) was used as a substitute for the perturbative Debye mass, effectively discarding the nonperturbative contributions. The perturbative quark mass does not suffer from this issue, and as such there is bound to be some ``mixing'' of prescriptions for the mass parameters of the NNLO full QCD results unless one accepts the variational complex $m_D$ value and sticks to the variational prescription for both parameters. In this section we will discuss and compare numerically the aforementioned prescriptions, before making a choice of prescriptions to use for comparison of our results to lattice data.

\subsection{Variational masses}

\begin{figure}[t]
\centering
\includegraphics[width=0.6\textwidth]{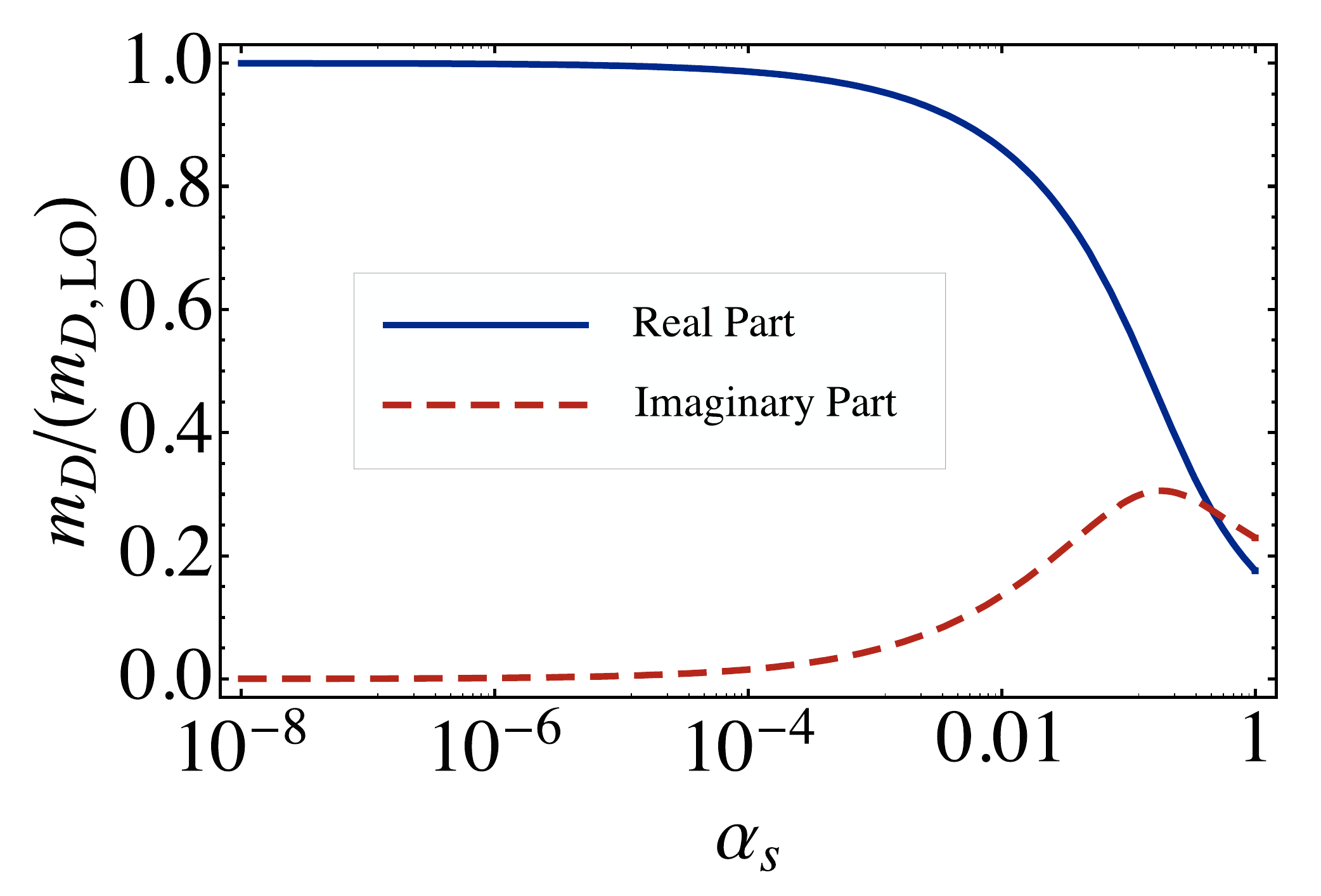}
\caption{Comparison of the real and imaginary valued contributions to the NNLO variational Debye mass, scaled by the LO perturbative Debye mass, with $N_c=3$ and $N_f=3$. Note that in the small coupling limit the variational and LO perturbative Debye masses converge.}
\label{fig:mdnnlogap-nf3}
\end{figure}

Solving the NNLO gap equations~(\ref{gapmd})-(\ref{gapmq}) yields a complex Debye mass and $m_q=0$. Figure~\ref{fig:mdnnlogap-nf3} shows the real and imaginary parts of the complex NNLO Debye mass, scaled by the LO perturbative $m_D$. While the imaginary contribution is non-negligible at intermediate coupling, and even grows to surpass the real contribution as $\alpha_s$ approaches unity, it disappears in the small coupling limit, where the NNLO variational and LO perturbative results for $m_D$ overlap. Presumably the complex mass parameter is an artifact of the truncation after the $m_D/T$ and $m_q/T$ expansions, but this is difficult to confirm without extending the truncation to higher order.

\begin{figure}[t]\centering
\includegraphics[width=0.6\textwidth]{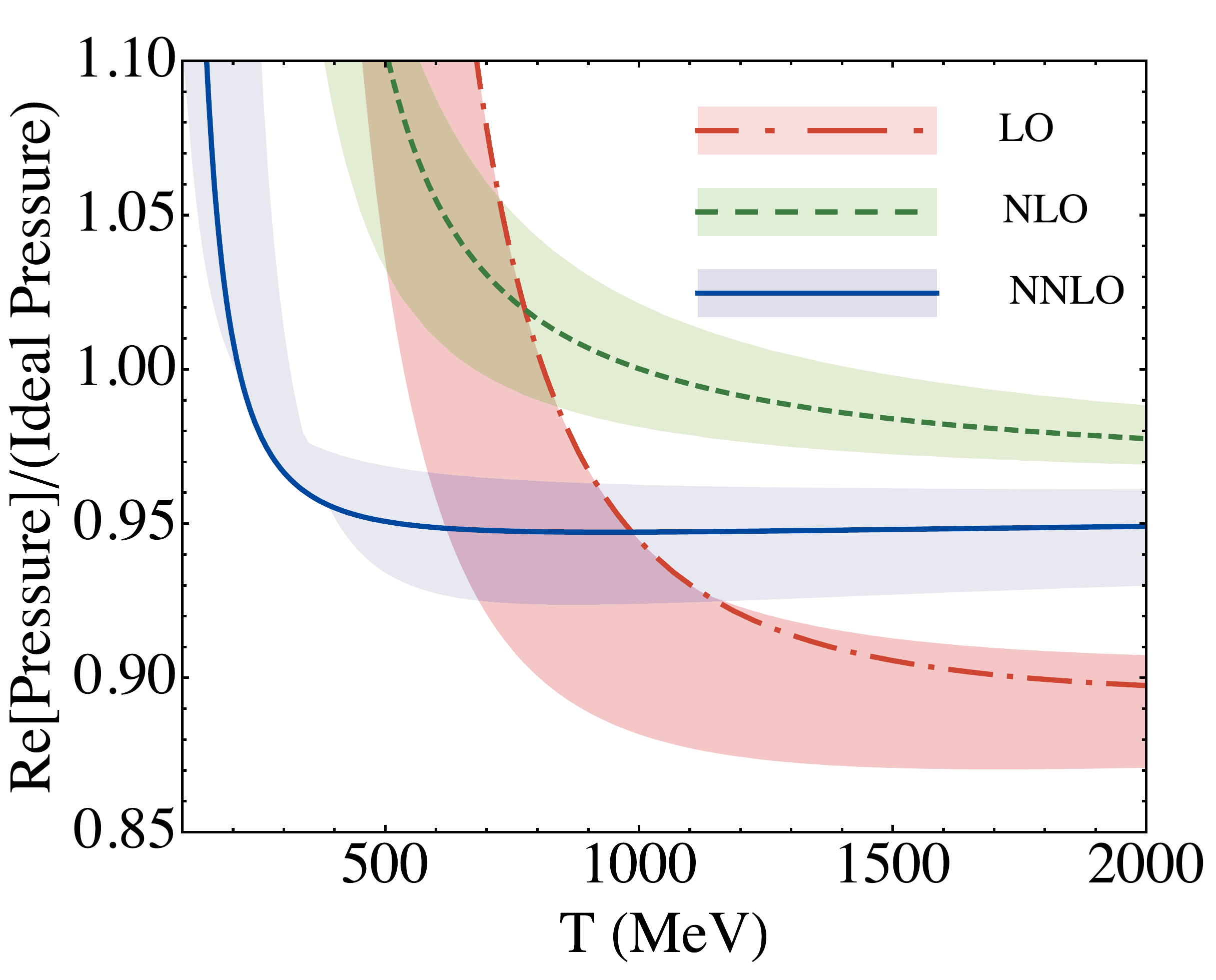}
\caption{The LO, NLO, and NNLO scaled variational pressures, with the imaginary contributions to the NNLO pressure discarded, $N_c=3$ and $N_f=3$. The shaded bands show the result of varying the renormalization scale $\mu$ by a factor of 2 around the central value $\mu=2\pi T$.}
\label{pressure-nf3-varmass}
\end{figure}

\begin{figure}[t]\centering
\includegraphics[width=0.6\textwidth]{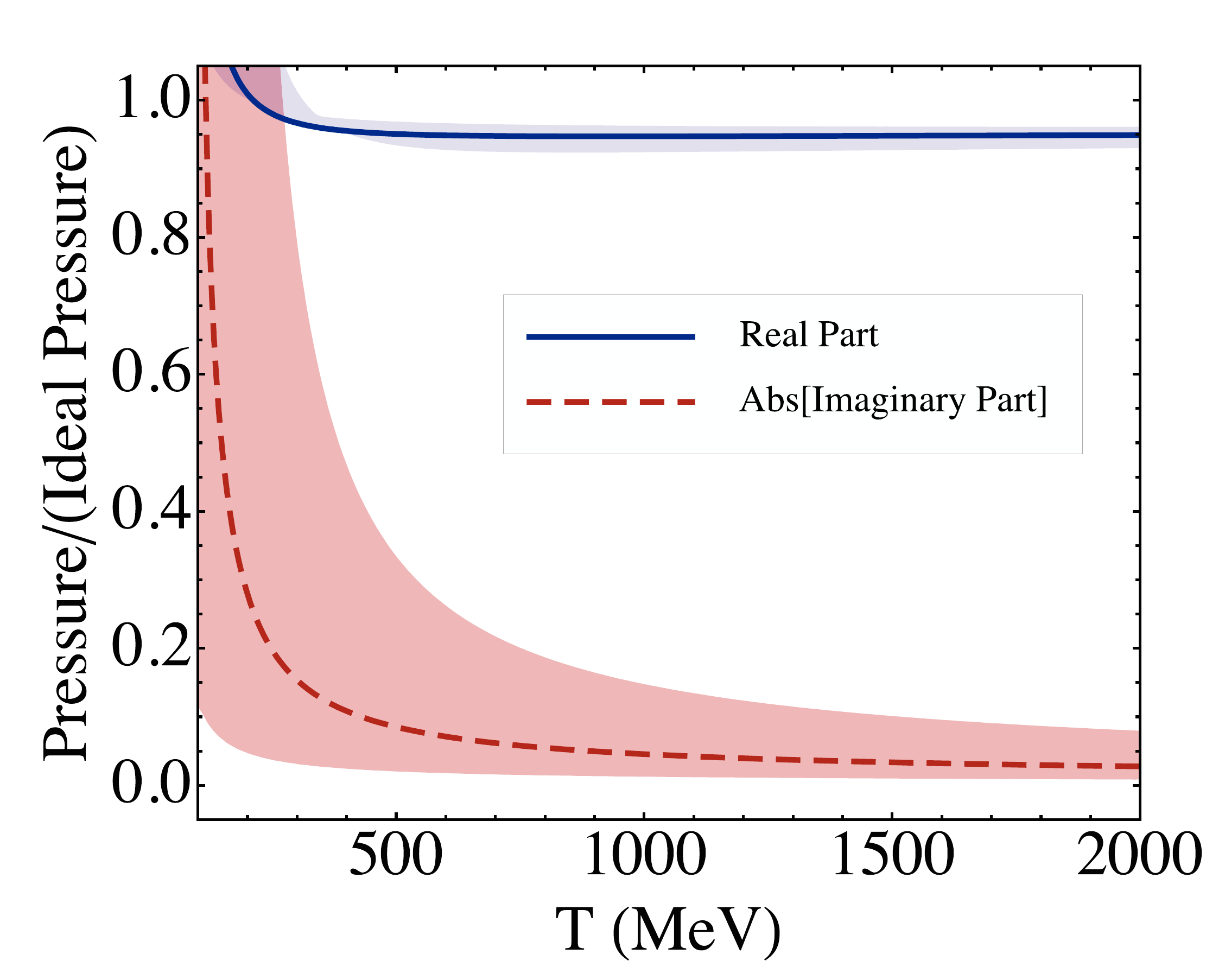}
\caption{Comparison of the real and imaginary contributions to the NNLO variational pressure, with $N_c=3$ and $N_f=3$. The shaded bands show the result of varying the renormalization scale $\mu$ by a factor of 2 around the central value $\mu=2\pi T$.}
\label{pressure-nf3-nnlovarmass-reim}
\end{figure}

 One strategy is then to throw away the imaginary part of the thermodynamic potential to obtain thermodynamic functions that are real valued. Figure~\ref{pressure-nf3-varmass} shows the NNLO pressure found using this approach. For reference, figure~\ref{pressure-nf3-nnlovarmass-reim} compares the real valued and the discarded imaginary contributions to the NNLO variational pressure over the same temperature range.

\subsection{Perturbative masses}

At LO in the coupling constant $g$, the Debye mass is given by the static longitudinal gluon self-energy at zero three-momentum, $m_D^2=\Pi_L(0,0)$, i.e.
\bqa
m^2_{D,LO} &=& g^2\left[c_A(d-1)^2\sumint_P{1\over P^2}-4s_F(d-1)\sumint_{\{P\}}{1\over P^2}\right]  \nonumber\\
&=& {4\pi\over3}\alpha_s T^2(c_A+s_F)\;.
\eqa
At NLO the weak-coupling perturbative Debye mass becomes logarithmically infrared divergent, reflecting the contribution from the nonperturbative magnetic scale $g^2T$~\cite{Linde:1980ts,Gross:1980br}. The nonperturbative contribution was calculated in ref.~\cite{Rebhan:1993az} and reads 
\bqa
\delta m_D^2 = m_D^2\sqrt{3c_A\over\pi}\alpha_s^{1/2}\left[\log{2m_D\over m_\textrm{mag}}-{1\over2}\right]\;,
\eqa
where $m_\textrm{mag}$ is the nonperturbative magnetic mass. Using this mass prescription for the Debye mass would then require input from e.g.~lattice simulations. For this reason we will not consider this prescription and instead will use the Braaten-Nieto prescription detailed in the next subsection.

The NLO quark mass is infrared safe, and was calculated for $N_c=3$ and $N_f=2$ in ref.~\cite{Carrington:2008dw},
\bqa
m_q = {gT \over \sqrt{6}}\left[1+(1.867\pm0.02){g \over 4\pi}\right] \;.
\eqa
However, to our knowledge there has not been results available for general $N_f$, and in order to make the comparison with lattice data feasible, we use instead, whenever it applies, the $N_c=3$ LO quark mass
\bqa
m_q = {gT \over \sqrt{6}} \;.
\eqa

\subsection{BN mass}

The strategy of equating $m_D$ to the BN mass has earlier been applied to NNLO HTLpt Yang-Mills theory \cite{Andersen:2009tc,Andersen:2010ct}. Inspired by dimensional reduction, one equates the $m_D$ mass parameter with the mass parameter of three-dimensional EQCD \cite{Braaten:1995ju}, i.e. $m_D=m_E$. This mass can be interpreted as the contribution to the Debye mass from the hard scale $T$ and is well defined and gauge invariant order-by-order in perturbation theory. However, beyond NLO it will also depend on the factorization scale that separates the hard scale and the soft scale $gT$.
In ref. \cite{Braaten:1995ju} it was calculated to NLO, giving
\bqa
m_D^2&=&
{4\pi\alpha_s\over3}T^2
\left\{
	c_A+s_F
	+{c_A^2\alpha_s\over3\pi}
		\left(
			{5\over 4}+{11\over 2}\gamma_E+{11\over 2}\log{\hat{\mu}\over 2}	
		\right)
	+{c_A s_F \alpha_s\over\pi}
		\left[
			{3\over 4}-{4\over 3}\log 2
			\right.\right.\nonumber\\&&\left.\left.
			+{7\over 6}\left(\gamma_E+\log{\hat{\mu}\over 2}\right)
		\right]
	+{s_F^2\alpha_s\over\pi}\left({1\over 3}-{4\over 3}\log 2-{2\over 3}\gamma_E-{2\over 3}\log{\hat{\mu}\over 2}\right)
	-{3\over 2}{s_{2F}\alpha_s\over\pi}
\right\}.
\nonumber\\
\label{eq:PmD}
\eqa

\subsection{Choice of mass prescriptions}

To avoid dealing with imaginary contributions to the thermodynamic potential, we opt for the BN prescription for the Debye mass parameter, the same choice that was made for NNLO HTLpt pure-glue QCD \cite{Andersen:2010ct}.

\begin{figure}[t]
\begin{center}
\includegraphics[width=0.49\textwidth]{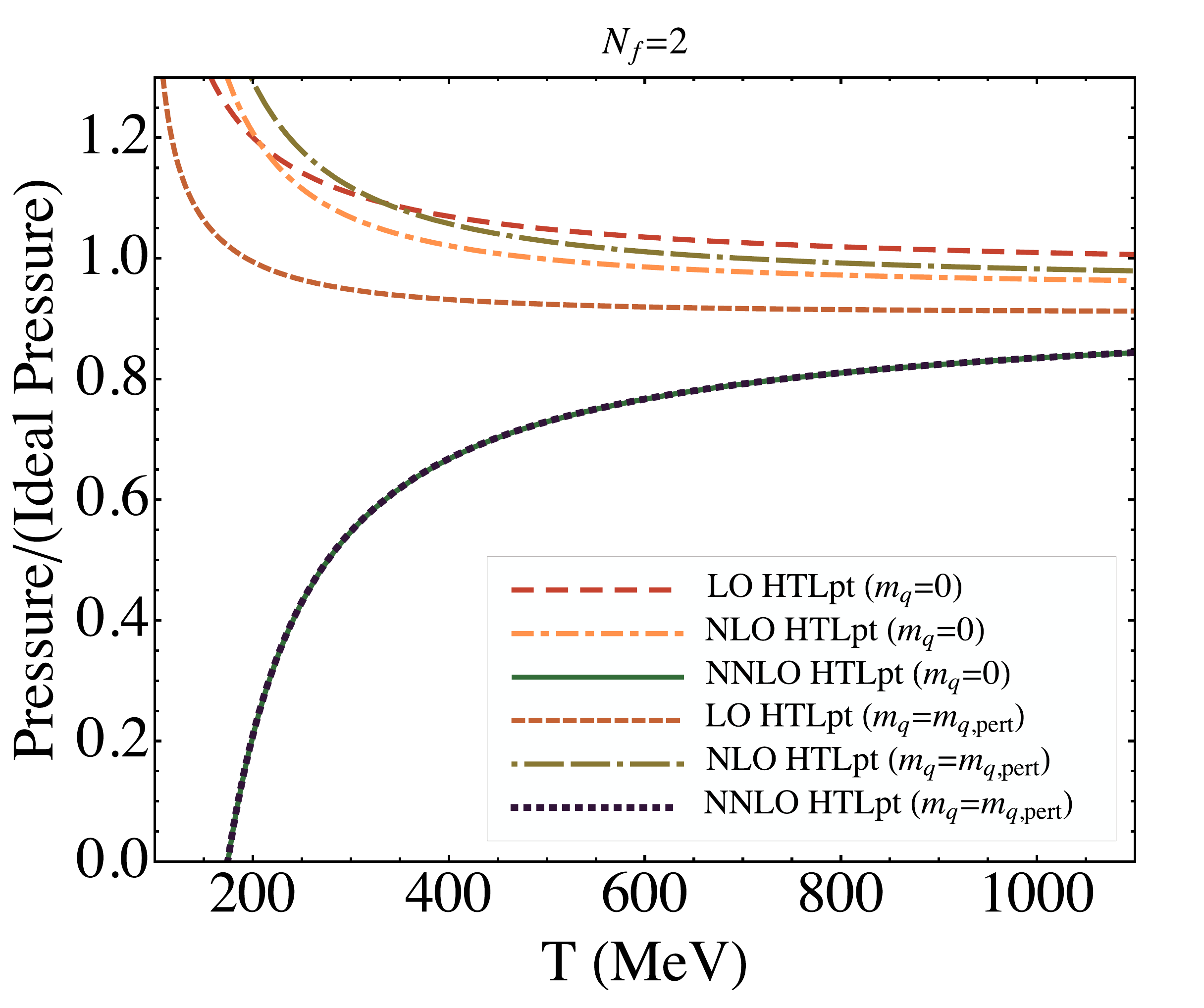}
\hspace{0.003\textwidth}
\includegraphics[width=0.49\textwidth]{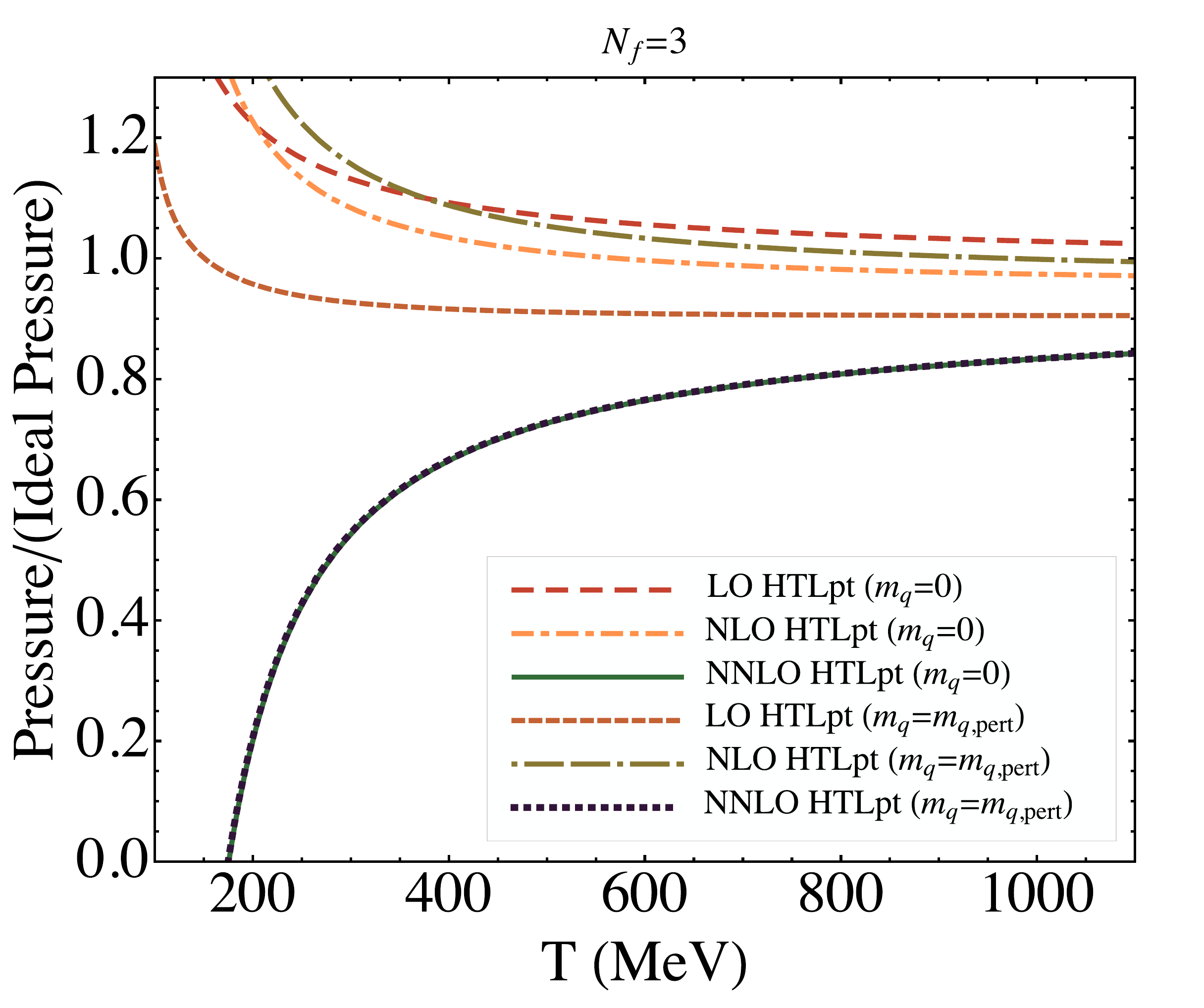}
\caption{Comparison of LO, NLO, and NNLO predictions for the scaled pressure for $N_c=3$, $N_f=2$ (left panel) and $N_f=3$ (right panel), using BN $m_D$ and both LO perturbative $m_q$ and $m_q=0$.}
\label{pressure-quarkmass-comp}
\end{center}
\end{figure}

\begin{figure}[t]
\begin{center}
\includegraphics[width=0.49\textwidth]{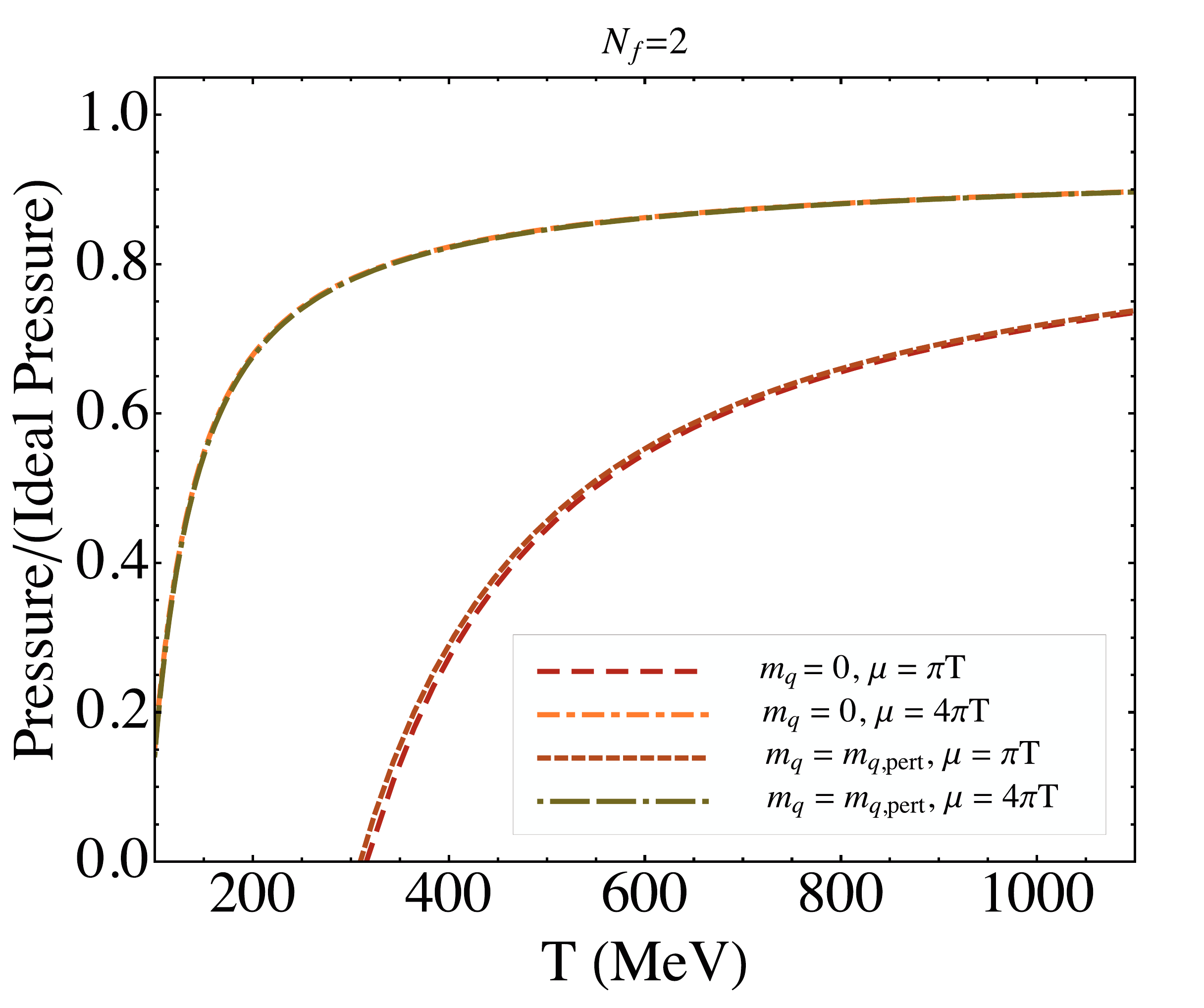}
\hspace{0.003\textwidth}
\includegraphics[width=0.49\textwidth]{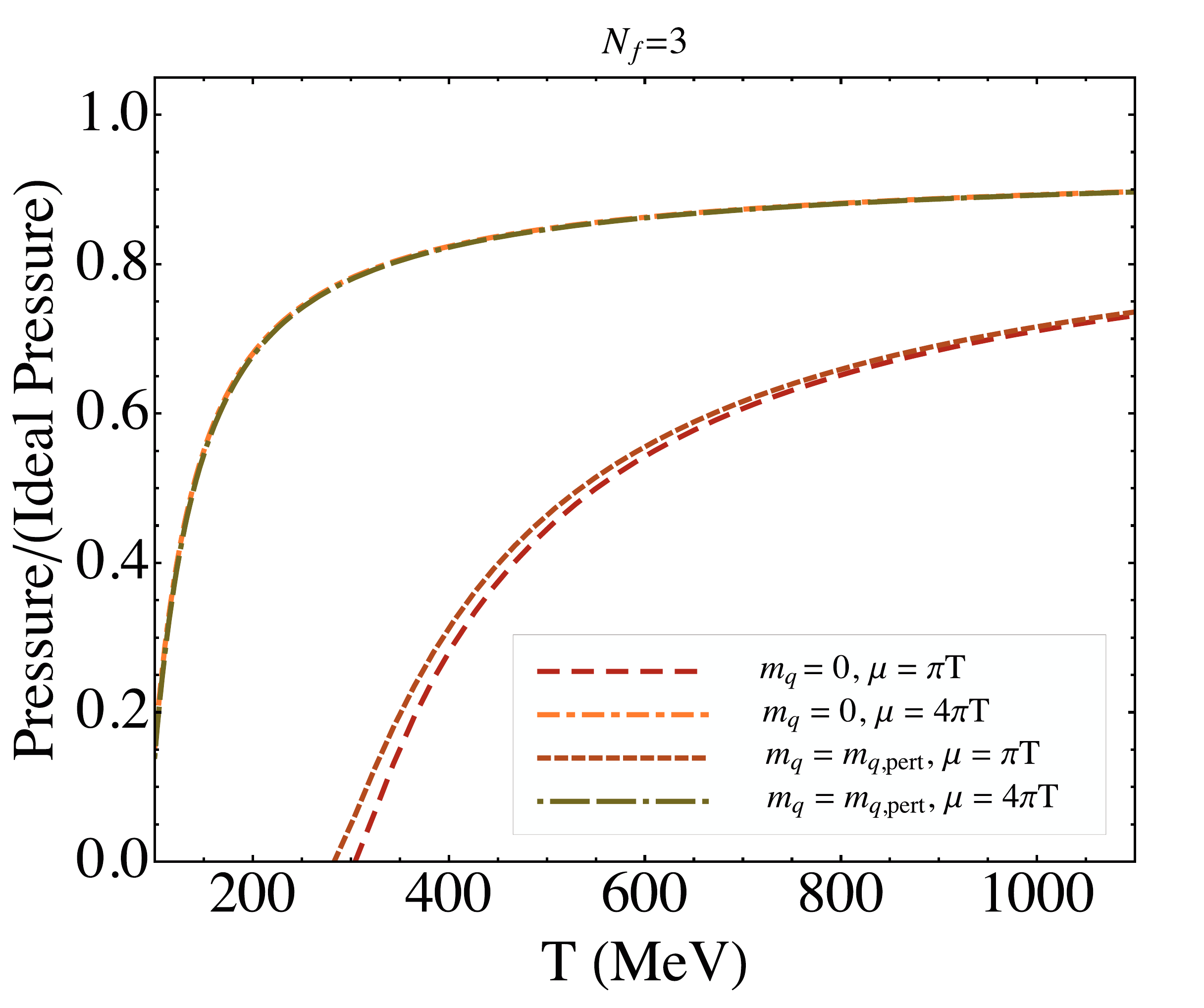}
\end{center}
\caption{NNLO predictions for the scaled pressure with BN $m_D$ comparing LO perturbative $m_q$ and $m_q=0$ at renormalization scales $\mu=4\pi T$ and $\mu=\pi T$, for $N_c=3$,  $N_f=2$ (left panel) and $N_f=3$ (right panel).}  
\label{pressureComp2}
\end{figure}

As for $m_q$, there is no problem with getting zero thermal quark mass from the variational prescription: even if the $m_q^2$-terms of Eq.~(\ref{Omega-NNLO}) drop out, the quarks still contribute to the HTLpt thermodynamical potential through the terms proportional to $d_F$, $s_F$ and $s_{2F}$. On the other hand, when using the BN mass parameter one might want to use the perturbative approach for the quark mass parameter, as using BN mass and perturbative thermal quark mass can be argued to be less of a ``mixing'' of prescriptions than BN mass and $m_q=0$. As it turns out, the final NNLO results are very insensitive to whether one chooses a perturbative mass prescription for $m_q$,  or uses the variational mass $m_q=0$. In figure~\ref{pressure-quarkmass-comp} the NNLO pressure with LO perturbative $m_q$ is virtually indistinguishable from the $m_q=0$ result. The difference in dependence on the renormalization scale is also very small, though slightly in favor of the perturbative $m_q$, see figure~\ref{pressureComp2}. However, convergence is improved with the choice $m_q=0$, where the pressure curves of figure~\ref{pressure-quarkmass-comp} monotonically approaching the NNLO results rather than oscillating as one goes from LO to NLO to NNLO. We will therefore use $m_q=0$ in the following unless otherwise stated.

\subsection{Comparison to lattice data}

\begin{figure}[t]
\begin{center}
\includegraphics[width=0.49\textwidth]{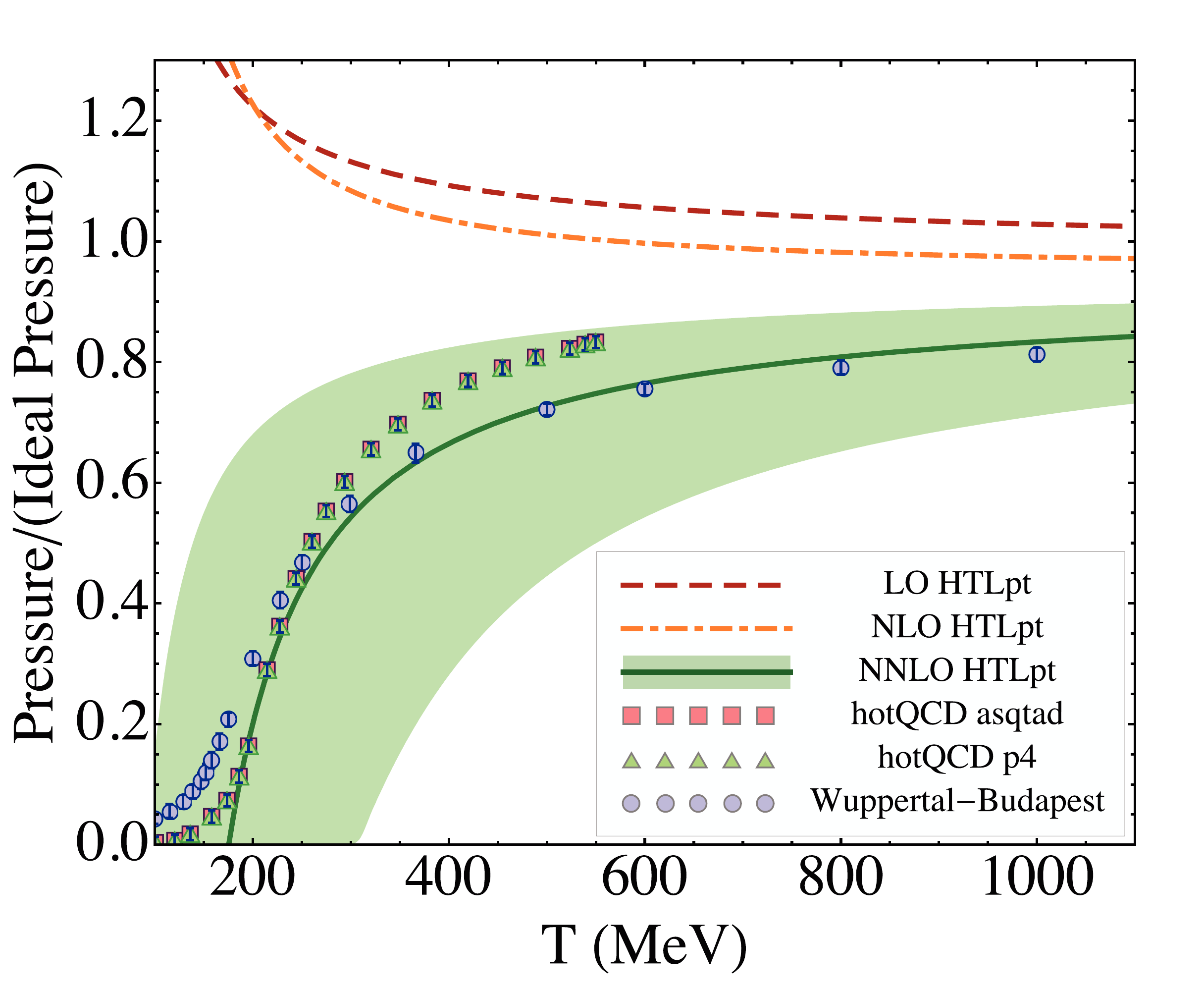}
\hspace{0.003\textwidth}
\includegraphics[width=0.49\textwidth]{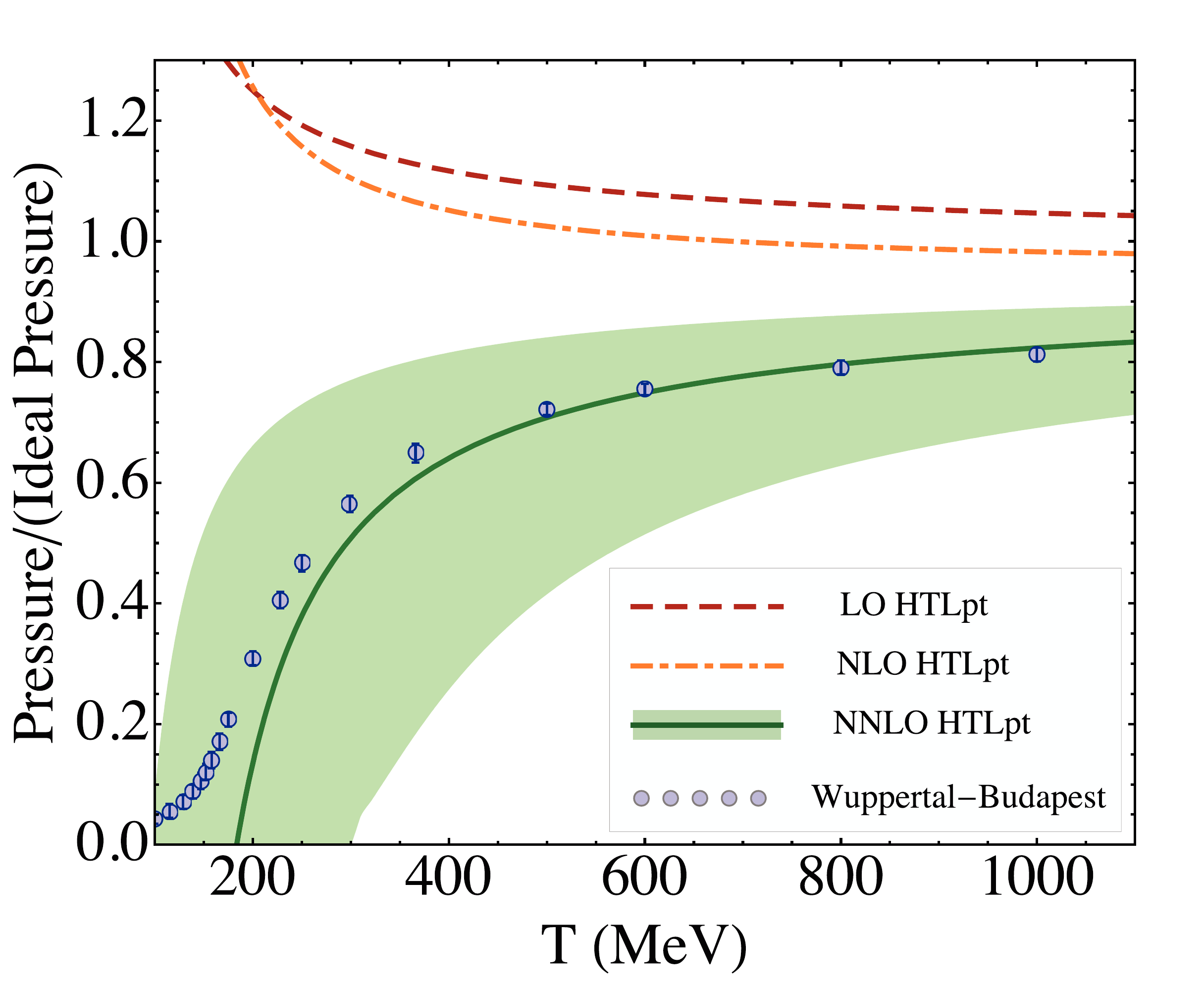}
\end{center}
\caption{Comparison of LO, NLO, and NNLO predictions for the scaled pressure for $N_f=2+1$ (left panel) and $N_f=2+1+1$ (right panel) with lattice data from Bazavov et al.~\cite{Bazavov:2009zn} and Borsanyi et al. \cite{Borsanyi:2010cj}. We use $N_c=3$, three-loop running for $\alpha_s$, $\mu=2\pi T$, and $\Lambda_{\overline{\rm MS}}=344\,$MeV. Shaded band shows the result of varying the renormalization scale $\mu$ by a factor of 2 around $\mu = 2 \pi T$ for the NNLO result. See main text for details.}  
\label{pressure}
\end{figure}

In figure~\ref{pressure} we show the normalized pressure for $N_c=3$ and $N_f=2+1$ (left panel), and $N_c=3$ and $N_f=2+1+1$ (right panel) as a function of $T$. The results at LO, NLO, and NNLO use the perturbative Debye mass given by eq.~(\ref{eq:PmD}) as well as $m_q=0$. For the strong coupling constant $\alpha_s$, we used three-loop running~\cite{Amsler:2008zzb} with $\Lambda_{\overline{\rm MS}}=344\,$MeV which for $N_f=3$ gives $\alpha_s({\rm 5\,GeV}) = 0.2034$~\cite{McNeile:2010ji}. The central line is evaluated with the renormalization scale $\mu = 2 \pi T$ which is the value one expects from effective field theory calculations \cite{Braaten:1995ju,Laine:2006cp} and the band represents a variation of $\mu$ by a factor of 2 around this value.

The lattice data from the Wuppertal-Budapest collaboration use the stout action. Since their results show essentially no dependence on the lattice spacing (it is smaller than the statistical errors), they provide a continuum estimate by averaging the trace anomaly measured using their two smallest lattice spacings corresponding to $N_\tau = 8$ and $N_\tau = 10$ \cite{Borsanyi:2010cj}, which were essentially on top of the $N_\tau=6$ measurement \cite{Aoki:2005vt}.\footnote{We note that the Wuppertal-Budapest group has published a few data points  for the trace anomaly with $N_\tau =12$ and within statistical error bars these are consistent with the published continuum estimated results.} Using standard lattice techniques, the continuum-estimated pressure is computed from an integral of the trace anomaly. The lattice data from the hotQCD collaboration are their $N_\tau = 8$ results using both the asqtad and p4 actions~\cite{Bazavov:2009zn}. The hotQCD results have not been continuum extrapolated and the error bars correspond to only statistical errors and do not factor in the systematic error associated with the calculation which, for the pressure, is estimated by the hotQCD collaboration to be between 5 - 10\%. We note that there are hotQCD results for physical light quark masses \cite{Cheng:2009zi}; however, these are available only for temperatures below 260\,MeV and the results are very close to the results shown in the figures so we do not include them here.  

As can be seen from figure~\ref{pressure} the successive HTLpt approximations represent an improvement over the successive approximations coming from a weak-coupling expansion; however, as in the pure-glue case~\cite{Andersen:2009tc,Andersen:2010ct}, the NNLO result represents a significant correction to the LO and NLO results. That being said, the NNLO HTLpt result agrees quite well with the available lattice data down to temperatures  on the order or $2\,T_c \sim 340\,$MeV for both $N_f=3$ (left panel) and $N_f=4$ (right panel).  Below these temperatures the successive approximations give large corrections with the correction from NLO to NNLO reaching 100\% near $T_c$.

\begin{figure}[t]
\begin{center}
\includegraphics[width=0.49\textwidth]{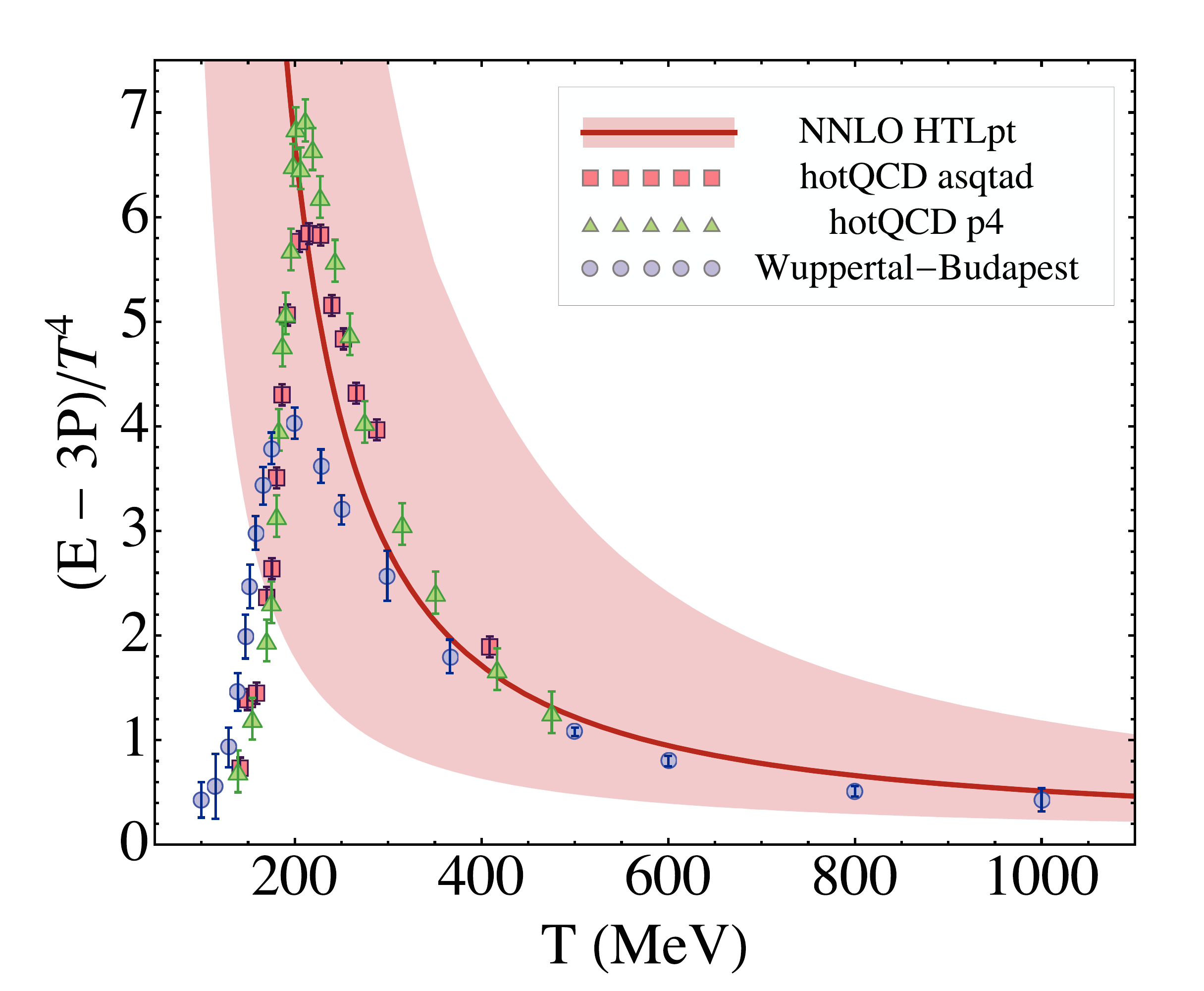}
\hspace{0.003\textwidth}
\includegraphics[width=0.49\textwidth]{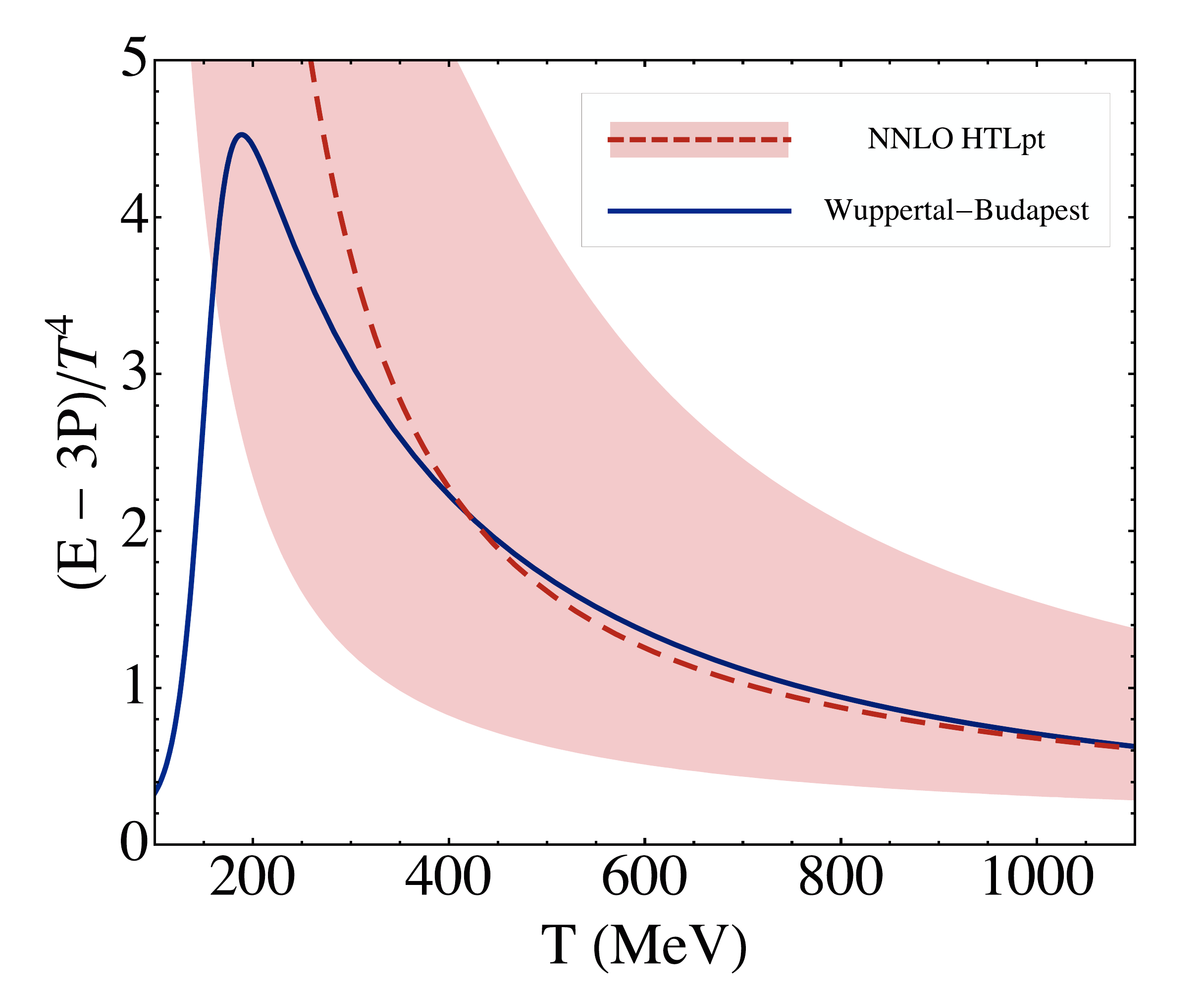}
\end{center}
\caption{Comparison of NNLO predictions for the scaled trace anomaly with $N_f=2+1$ (left panel) and $N_f=2+1+1$ fermions (right panel) lattice data from Bazavov et al.~\cite{Bazavov:2009zn} and Borsanyi et al. \cite{Borsanyi:2010cj}. We use $N_c=3$, three-loop running for $\alpha_s$, $\mu=2\pi T$, and $\Lambda_{\overline{\rm MS}}=344\,$MeV. Shaded band shows the result of varying the renormalization scale $\mu$ by a factor of 2 around $\mu = 2 \pi T$. See main text for details.}
\label{trace2}
\end{figure}

In figure~\ref{trace2}, we show the NNLO approximation to the trace anomaly (interaction measure) normalized to $T^4$ as a function of $T$ for $N_c=3$ and $N_f=2+1$ (left panel) and for $N_c=3$ and $N_f=2+1+1$ (right panel).\footnote{The Wuppertal-Budapest data were obtained using a physical charm mass at low temperatures; however, we use massless quarks. The difference between massive and massless quarks is expected to be significant only for $T \lsim 414\,$MeV corresponding to the temperature where the lowest fermionic Matsubara mode equals the charm quark mass.} In the left panel we show data from both the Wuppertal-Budapest collaboration and the hotQCD collaboration taken from the same data sets displayed in figure~\ref{pressure} and described previously. In the case of the hotQCD results we note that the results for the trace anomaly using the p4 action show large lattice size affects at all temperatures shown and the asqtad results for the trace anomaly show large lattice size effects for $T \gsim 200\,$MeV. In the right panel we display a parameterization (solid blue curve) of the trace anomaly for $N_f=4$ published by the Wuppertal-Budapest collaboration \cite{Borsanyi:2010cj} since the individual data points were not published. In both the left and right panels we see very good agreement with the available lattice data down to temperatures on the order of $T \sim 2\,T_c$.

\subsection{Large $N_f$}

In the limit $N_f \gg 1$ while keeping $g^2 N_f \sim 1$, only ring diagrams contribute to the pressure in perturbation theory. Since $s_F$ is proportional to $N_f$ in both QCD and QED, this indicates the equivalence of QED and QCD in the large-$N_f$ limit, with the large $N_f$ effective coupling defined as $g_{\rm eff} \equiv g\sqrt{N_f/2}$ for QCD, and $g_{\rm eff} \equiv e\sqrt{N_f}$ for QED. The NNLO HTLpt results in this limit were discussed in the context of QED in ref.~\cite{Su:2011zv}. 

\begin{figure}[t]
\centering
	\includegraphics[width=0.6\textwidth]{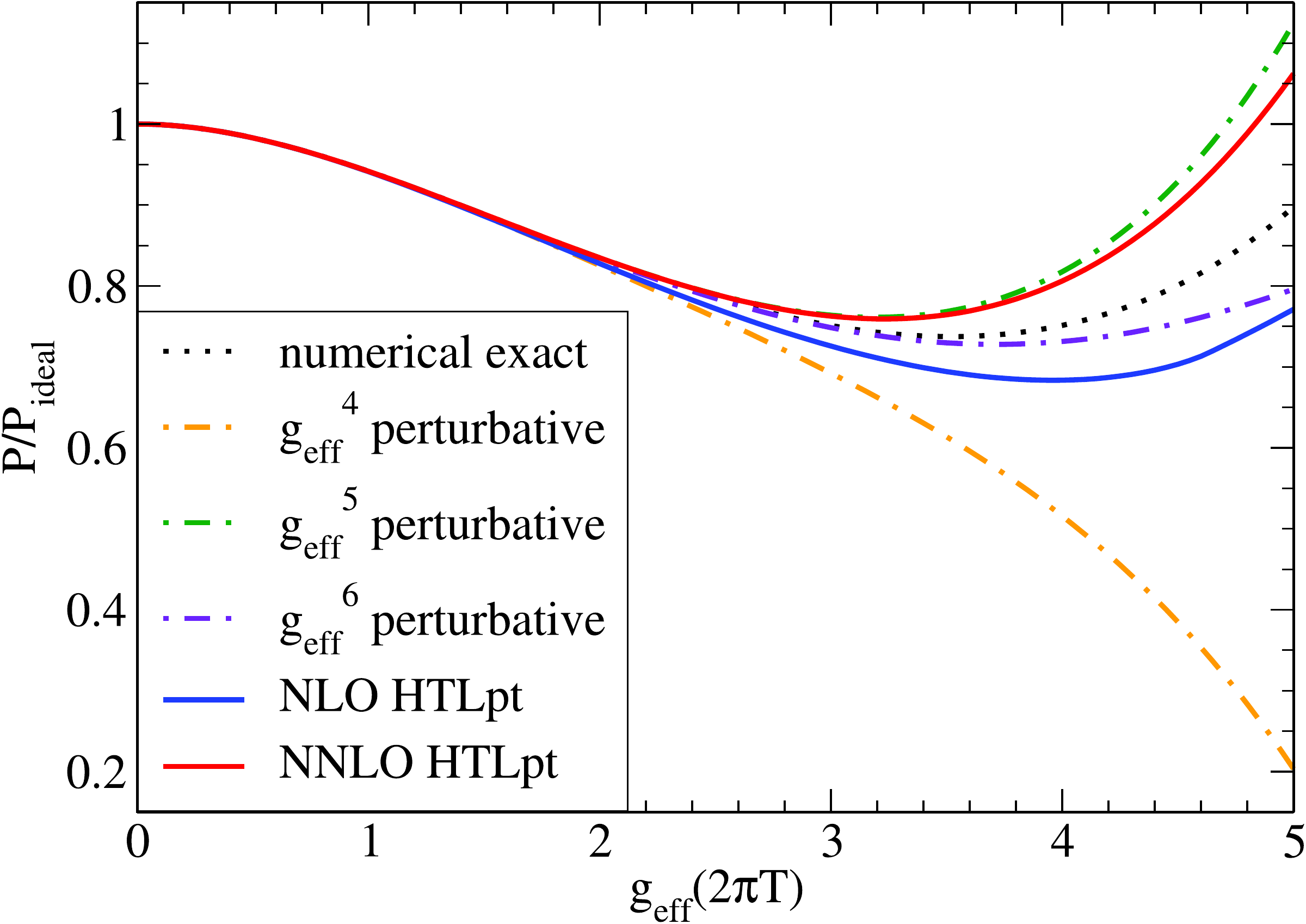}
\caption{The predictions for the large-$N_f$ pressure of QCD between the numerical exact result from \cite{Ipp:2003zr}, NLO and NNLO HTLpt, and perturbative $g_\mathrm{eff}^4$ through $g_\mathrm{eff}^6$ \cite{Gynther:2009qf} results at $\mu=e^{-\gamma_E}\pi T$.}
\label{fig:LargeNf}
\end{figure}

It is possible to solve for the $O(N_f^0)$ contribution to the pressure exactly \cite{Ipp:2003zr}, enabling comparison between predictions from approximations such as perturbation theory or HTLpt, and exact numerical results. In figure~\ref{fig:LargeNf}, we plot the NLO and NNLO HTLpt predictions for the large-$N_f$ pressure along with the numerical result of \cite{Ipp:2003zr} as well as the perturbative $g^4_\mathrm{eff}$, $g^5_\mathrm{eff}$, and $g^6_\mathrm{eff}$ predictions at $\mu =e^{-\gamma_E}\pi T$ obtained in \cite{Gynther:2009qf}. For the pressure, going to NNLO in HTLpt extends the range of agreement with exact results compared to NLO, from $g_\mathrm{eff}\lesssim 2$ to $g_\mathrm{eff}\lesssim 2.8$. The HTLpt large coupling behavior qualitatively matches that of the numerical results.

\section{Summary and outlook}
\label{sec:7}

We have calculated the NNLO contributions to the thermodynamic functions of SU($N_c$) Yang-Mills theory with $N_f$ fermions using HTLpt. Using the BN mass for $m_D$ and $m_q=0$ as the mass prescriptions, at NNLO we find that HTLpt predictions for the pressure and the trace anomaly are in agreement with lattice data for $N_c=3$ and $N_f \in \{3,4\}$ down to $T\sim2\,T_c$. The failure of HTLpt to match lattice data at lower temperatures is to be expected since one is expanding around the trivial vacuum $A_{\mu}=0$ and therefore neglects the approximate center symmetry $Z(N_c)$, which becomes essential as one approaches the deconfinement transition~\cite{KorthalsAltes:1999xb,KorthalsAltes:2000gs,Pisarski:2000eq,Vuorinen:2006nz,deForcrand:2008aw,Hidaka:2009hs,Dumitru:2010mj}. In addition, it is also in line with expectations since below $T\sim2-3\,T_c$ a simple ``electric'' quasiparticle approximation breaks down due to nonperturbative chromomagnetic effects~~\cite{Linde:1980ts,Gross:1980br}.

We find that when including quarks the agreement with lattice data is greatly improved as compared to the NNLO results of pure-glue QCD~\cite{Andersen:2009tc,Andersen:2010ct}. Fermions are perturbative in the sense that they decouple in the dimensional-reduction step of effective field theory, so we expect that including contributions from quarks gives at least as good agreement with the lattice calculations as the pure-glue case. However, the exact reason for the better agreement between the HTLpt predictions and lattice calculations when including quarks is not clear to us.

Just as for NNLO pure-glue QCD we found that the variational solution for the mass parameter $m_D$ is complex and we therefore chose instead to use the Debye mass from EQCD for our $m_D$ parameter. In the variational prescription the quark gap equation gives $m_q=0$, but even if the quarks do not get a thermal mass, the thermodynamical potential still gets HTLpt contributions from the inclusion of quarks in the calculation. Whether the complexity of the variational Debye mass parameter is due to the additional expansion in $m_D/T$ and $m_q/T$ is impossible to decide at this stage. The correction to the pressure going from NLO to NNLO is also rather large. It is unfortunate that the nonperturbative magnetic scale prevents going to N$^3$LO without supplementing the calculation with input from three-dimensional lattice calculations, as it would be interesting to see whether the complexity of the Debye mass parameter and the slow convergence persists.

In closing, we emphasize that HTLpt provides a gauge invariant reorganization of perturbation theory for calculating static and dynamic quantities in thermal field theory. Given the good agreement with lattice data for thermodynamic quantities down to temperatures that are relevant for LHC, it would therefore be interesting and challenging to apply HTLpt to the calculation of dynamic quantities, especially transport coefficients, at these temperatures.

\section*{Acknowledgments}

The authors would like to thank S.~Borsanyi for providing us with the latest lattice data of the Wuppertal-Budapest collaboration. We thank S.~Borsanyi and Z.~Fodor for useful discussions. N.~Su thanks the Department of Physics at the Norwegian University of Science and Technology (NTNU) and the Kavli Institute for Theoretical Physics China (KITPC) for kind hospitality. M.~Strickland was supported by the Helmholtz International Center for FAIR within the framework of the LOEWE program launched by the State of Hesse. N.~Su was support by the Helmholtz International Center for FAIR within the framework of the LOEWE program launched by the State of Hesse and the Research Council of Norway through the Yggdrasil mobility programme.

\bibliography{htlpt}

\providecommand{\href}[2]{#2}\begingroup\raggedright\begin{thebibliography}{10}

\bibitem{Aoki:2006we}
  Y.~Aoki, G.~Endrodi, Z.~Fodor, S.~D.~Katz and K.~K.~Szabo, 
  {\it {The order of the quantum chromodynamics transition predicted by the 
  standard model of particle physics}},  
  {\em Nature} {\bf 443} (2006) 675,
  [\href{http://xxx.lanl.gov/abs/hep-lat/0611014}{{\tt hep-lat/0611014}}].

\bibitem{Shuryak:1977ut}
  E.~V.~Shuryak, 
  {\it {Theory of hadron plasma}},  
  {\em Sov. Phys. JETP} {\bf 47} (1978) 212.

\bibitem{Kapusta:1979fh}
  J.~I.~Kapusta, 
  {\it {Quantum chromodynamics at high temperature}},  
  {\em Nucl. Phys.} {\bf B 148} (1979) 461.

\bibitem{Toimela:1984xy}
  T.~Toimela, 
  {\it {Perturbative QED and QCD at finite temperatures and densities}},  
  {\em Int. J. Theor. Phys.} {\bf 24} (1985) 901.

\bibitem{Arnold:1994ps}
  P.~B.~Arnold and C.-X.~Zhai, 
  {\it {Three-loop free energy for pure gauge QCD}},  
  {\em Phys. Rev.} {\bf D 50} (1994) 7603,
  [\href{http://xxx.lanl.gov/abs/hep-ph/9408276}{{\tt hep-ph/9408276}}].

\bibitem{Arnold:1994eb}
  P.~B.~Arnold and C.-X.~Zhai, 
  {\it {Three-loop free energy for high-temperature QED and QCD with fermions}},  
  {\em Phys. Rev.} {\bf D 51} (1995) 1906, 
  [\href{http://xxx.lanl.gov/abs/hep-ph/9410360}{{\tt hep-ph/9410360}}].

\bibitem{Braaten:1995ju}
  E.~Braaten and A.~Nieto, 
  {\it {On the Convergence of Perturbative QCD at High Temperature}},  
  {\em Phys. Rev. Lett.} {\bf 76} (1996) 1417,
  [\href{http://xxx.lanl.gov/abs/hep-ph/9508406}{{\tt hep-ph/9508406}}].

\bibitem{Braaten:1995jr}
  E.~Braaten and A.~Nieto, 
  {\it {Free energy of QCD at high temperature}},  
  {\em Phys. Rev.} {\bf D 53} (1996) 3421,
  [\href{http://xxx.lanl.gov/abs/hep-ph/9510408}{{\tt hep-ph/9510408}}].

\bibitem{Zhai:1995ac}
  C.-X.~Zhai and B.~M.~Kastening, 
  {\it {Free energy of hot gauge theories with fermions through $g^5$}},  
  {\em Phys. Rev.} {\bf D 52} (1995) 7232,
  [\href{http://xxx.lanl.gov/abs/hep-ph/9507380}{{\tt hep-ph/9507380}}].

\bibitem{Kajantie:2002wa}
  K.~Kajantie, M.~Laine, K.~Rummukainen and Y.~Schroder, 
  {\it {Pressure of hot QCD up to $g^6 \ln(1/g)$}},  
  {\em Phys. Rev.} {\bf D 67} (2003) 105008,
  [\href{http://xxx.lanl.gov/abs/hep-ph/0211321}{{\tt hep-ph/0211321}}].
  
\bibitem{Amsler:2008zzb}
  Particle Data Group Collaboration, C.~Amsler {\em et.~al.}, 
  {\it {Review of Particle Physics}},  
  {\em Phys. Lett.} {\bf B 667} (2008) 1.
  
\bibitem{McNeile:2010ji}
  C.~McNeile, C.~T.~H.~Davies, E.~Follana, K.~Hornbostel and G.~P.~Lepage, 
  {\it {High-precision $c$ and $b$ Masses, and QCD coupling from current-current
  correlators in lattice and continuum QCD}},  
  {\em Phys. Rev.} {\bf D 82} (2010) 034512, 
  [\href{http://xxx.lanl.gov/abs/1004.4285}{{\tt arXiv:1004.4285}}].

\bibitem{Arsene:2004fa}
  BRAHMS Collaboration, I.~Arsene {\em et.~al.}, 
  {\it {Quark-gluon plasma and color glass condensate at RHIC? 
  The perspective from the BRAHMS experiment}},  
  {\em Nucl. Phys.} {\bf A 757} (2005) 1,
  [\href{http://xxx.lanl.gov/abs/nucl-ex/0410020}{{\tt nucl-ex/0410020}}].

\bibitem{Back:2004je}
  PHOBOS Collaboration, B.~B.~Back {\em et.~al.}, 
  {\it {The PHOBOS perspective on discoveries at RHIC}},  
  {\em Nucl. Phys.} {\bf A 757} (2005) 28,
  [\href{http://xxx.lanl.gov/abs/nucl-ex/0410022}{{\tt nucl-ex/0410022}}].

\bibitem{Adams:2005dq}
  STAR Collaboration, J.~Adams {\em et.~al.}, 
  {\it {Experimental and theoretical challenges in the search for the quark-gluon plasma: 
  The STAR Collaboration's critical assessment of the evidence from RHIC collisions}},
  {\em Nucl. Phys.} {\bf A 757} (2005) 102,
  [\href{http://xxx.lanl.gov/abs/nucl-ex/0501009}{{\tt nucl-ex/0501009}}].

\bibitem{Adcox:2004mh}
  PHENIX Collaboration, K.~Adcox {\em et.~al.}, 
  {\it {Formation of dense partonic matter in relativistic nucleus nucleus collisions at RHIC:
  Experimental evaluation by the PHENIX Collaboration}},  
  {\em Nucl. Phys.} {\bf A 757} (2005) 184,
  [\href{http://xxx.lanl.gov/abs/nucl-ex/0410003}{{\tt nucl-ex/0410003}}].

\bibitem{Gyulassy:2004zy}
  M.~Gyulassy and L.~McLerran, 
  {\it {New forms of QCD matter discovered at RHIC}},  
  {\em Nucl. Phys.} {\bf A 750} (2005) 30,
  [\href{http://xxx.lanl.gov/abs/nucl-th/0405013}{{\tt nucl-th/0405013}}].

\bibitem{Maldacena:1997re}
  J.~M.~Maldacena, 
  {\it {The Large-$N$ Limit of Superconformal Field Theories and Supergravity}},  
  {\em Adv. Theor. Math. Phys.} {\bf 2} (1998) 231,
  [\href{http://xxx.lanl.gov/abs/hep-th/9711200}{{\tt hep-th/9711200}}].

\bibitem{Qin:2007rn}
  G.~-Y.~Qin, J.~Ruppert, C.~Gale, S.~Jeon, G.~D.~Moore and M.~G.~Mustafa,
  {\it {Radiative and Collisional Jet Energy Loss in the Quark-Gluon Plasma at the BNL Relativistic Heavy Ion Collider}},
  {\em Phys. Rev. Lett.} {\bf 100} (2008) 072301,
  [\href{http://xxx.lanl.gov/abs/0710.0605}{{\tt arXiv:0710.0605}}].
  
\bibitem{Qin:2009gw}
  G.~-Y.~Qin and A.~Majumder,
  {\it {Perturbative QCD Description of Heavy and Light Flavor Jet Quenching}},
  {\em Phys. Rev. Lett.} {\bf 105} (2010) 262301,
  [\href{http://xxx.lanl.gov/abs/0910.3016}{{\tt arXiv:0910.3016}}].
  
\bibitem{Xu:2007jv}
  Z.~Xu, C.~Greiner and H.~St{\"o}cker,
  {\it {Perturbative QCD Calculations of Elliptic Flow and Shear Viscosity in Au+Au Collisions at $\sqrt{s_{NN}}=200\,$GeV}},
  {\em Phys. Rev. Lett.} {\bf 101} (2008) 082302,
  [\href{http://xxx.lanl.gov/abs/0711.0961}{{\tt arXiv:0711.0961}}]. 
  
\bibitem{Bazavov:2009zn}
  A.~Bazavov {\em et.~al.}, 
  {\it {Equation of state and QCD transition at finite temperature}},  
  {\em Phys. Rev.} {\bf D 80} (2009) 014504,
  [\href{http://xxx.lanl.gov/abs/0903.4379}{{\tt arXiv:0903.4379}}].

\bibitem{Borsanyi:2010cj}
  S.~Borsanyi {\em et.~al.}, 
  {\it {The QCD equation of state with dynamical quarks}},  
  \href{http://xxx.lanl.gov/abs/1007.2580}{{\tt arXiv:1007.2580}}.

\bibitem{Blaizot:2003tw}
  J.-P.~Blaizot, E.~Iancu and A.~Rebhan,
   {\it {Thermodynamics of the high-temperature quark gluon plasma}},
  \href{http://xxx.lanl.gov/abs/hep-ph/0303185}{{\tt hep-ph/0303185}}.

\bibitem{Kraemmer:2003gd}
  U.~Kraemmer and A.~Rebhan, 
  {\it {Advances in perturbative thermal field theory}},  
  {\em Rept. Prog. Phys.} {\bf 67} (2004) 351,
  [\href{http://xxx.lanl.gov/abs/hep-ph/0310337}{{\tt hep-ph/0310337}}].

\bibitem{Andersen:2004fp}
  J.~O.~Andersen and M.~Strickland, 
  {\it {Resummation in hot field theories}},
  {\em Ann. Phys.} {\bf 317} (2005) 281,
  [\href{http://xxx.lanl.gov/abs/hep-ph/0404164}{{\tt hep-ph/0404164}}].

\bibitem{Karsch:1997gj}
  F.~Karsch, A.~Patkos and P.~Petreczky, 
  {\it {Screened perturbation theory}},
  {\em Phys. Lett.} {\bf B 401} (1997) 69,
  [\href{http://xxx.lanl.gov/abs/hep-ph/9702376}{{\tt hep-ph/9702376}}].
  
\bibitem{Chiku:1998kd}
  S.~Chiku and T.~Hatsuda,
  {\it {Optimized perturbation theory at finite temperature}},
  {\em Phys. Rev.}  {\bf D 58 } (1998) 076001,
  [\href{http://xxx.lanl.gov/abs/hep-ph/9803226}{{\tt hep-ph/9803226}}].

\bibitem{Andersen:2000yj}
  J.~O.~Andersen, E.~Braaten and M.~Strickland, 
  {\it {Screened perturbation theory to three loops}},  
  {\em Phys. Rev.} {\bf D 63} (2001) 105008,
  [\href{http://xxx.lanl.gov/abs/hep-ph/0007159}{{\tt hep-ph/0007159}}].
    
\bibitem{Andersen:2001ez}
  J.~.O.~Andersen and M.~Strickland,
  {\it {Mass expansions of screened perturbation theory}},
  {\em Phys. Rev.}  {\bf D 64} (2001) 105012,
  [\href{http://xxx.lanl.gov/abs/hep-ph/0105214}{{\tt hep-ph/0105214}}].  
  
\bibitem{Andersen:2008bz}
  J.~O.~Andersen and L.~Kyllingstad,
  {\it {Four-loop Screened Perturbation Theory}},
  {\em Phys. Rev.}  {\bf D 78} (2008) 076008,
  [\href{http://xxx.lanl.gov/abs/0805.4478}{{\tt arXiv:0805.4478}}].  
  
\bibitem{Yukalov:1976pm}
  V.~I.~Yukalov,
  {\it {Remarks on quasiaverages}},
  {\em Teor. Mat. Fiz.} {\bf 26} (1976) 403.
  
\bibitem{Stevenson:1981vj}
  P.~M.~Stevenson,
  {\it {Optimized perturbation theory}},
  {\em Phys. Rev.} {\bf D 23} (1981) 2916.  

\bibitem{Duncan:1988hw}
  A.~Duncan and M.~Moshe,
  {\it {Nonperturbative physics from interpolating actions}},
  {\em Phys. Lett.} {\bf B 215} (1988) 352.

\bibitem{Duncan:1992ba}
  A.~Duncan and H.~F.~Jones,
  {\it {Convergence proof for optimized $\delta$ expansion: Anharmonic oscillator}},
  {\em Phys. Rev.} {\bf D 47} (1993) 2560.
  
\bibitem{Sisakian:1994nn}
  A.~N.~Sisakian, I.~L.~Solovtsov and O.~Shevchenko,
  {\it {Variational perturbation theory}},
  {\em Int. J. Mod. Phys.} {\bf A 9 } (1994) 1929.
  
\bibitem{Janke:1995zz}
  W.~Janke and H.~Kleinert,
  {\it {Convergent Strong-Coupling Expansions from Divergent Weak-Coupling Perturbation Theory}},
  {\em Phys. Rev. Lett.}  {\bf 75 } (1995) 2787.
  
\bibitem{Braaten:1991gm}
  E.~Braaten and R.~D.~Pisarski,
  {\it {Simple effective Lagrangian for hard thermal loops}},
  {\em Phys. Rev.} {\bf D 45} (1992) R1827.

\bibitem{Buchmuller:1994qy}
  W.~Buchmuller and O.~Philipsen,
  {\it {Phase structure and phase transition of the $SU(2)$ Higgs model in three-dimensions}},
  {\em Nucl. Phys.} {\bf B 443} (1995) 47,
  [\href{http://xxx.lanl.gov/abs/hep-ph/9411334}{{\tt hep-ph/9411334}}].

\bibitem{Alexanian:1995rp}
  G.~Alexanian and V.~P.~Nair,
  {\it {A self-consistent inclusion of magnetic screening for the quark-gluon plasma}},
  {\em Phys. Lett.} {\bf B 352} (1995) 435,
  [\href{http://xxx.lanl.gov/abs/hep-ph/9504256}{{\tt hep-ph/9504256}}].
  
\bibitem{bp}
  E.~Braaten and R.~D.~Pisarski, 
  {\it {Soft amplitudes in hot gauge theories: A general analysis}},
  {\em Nucl. Phys.} {\bf B 337} (1990) 569.
  
\bibitem{Andersen:1999fw}
  J.~O.~Andersen, E.~Braaten and M.~Strickland, 
  {\it {Hard-Thermal-Loop Resummation of the Free Energy of a Hot Gluon Plasma}},  
  {\em Phys. Rev. Lett.} {\bf 83} (1999) 2139,
  [\href{http://xxx.lanl.gov/abs/hep-ph/9902327}{{\tt hep-ph/9902327}}].

\bibitem{Andersen:1999sf}
  J.~O.~Andersen, E.~Braaten and M.~Strickland,
  {\it {Hard-thermal-loop resummation of the thermodynamics of a hot gluon plasma}},
  {\em Phys. Rev.}  {\bf D 61} (2000) 014017,
  [\href{http://xxx.lanl.gov/abs/hep-ph/9905337}{{\tt hep-ph/9905337}}].

\bibitem{Andersen:1999va}
  J.~O.~Andersen, E.~Braaten and M.~Strickland, 
  {\it {Hard-thermal-loop resummation of the free energy of a hot quark-gluon plasma}},  
  {\em Phys. Rev.} {\bf D 61} (2000) 074016,
  [\href{http://xxx.lanl.gov/abs/hep-ph/9908323}{{\tt hep-ph/9908323}}].

\bibitem{Andersen:2009tw}
  J.~O.~Andersen, M.~Strickland and N.~Su, 
  {\it {Three-loop hard-thermal-loop free energy for QED}},  
  {\em Phys. Rev.} {\bf D 80} (2009) 085015,
  [\href{http://xxx.lanl.gov/abs/0906.2936}{{\tt arXiv:0906.2936}}].

\bibitem{Andersen:2002ey}
  J.~O.~Andersen, E.~Braaten, E.~Petitgirard and M.~Strickland, 
  {\it {Hard-thermal-loop perturbation theory to two loops}},  
  {\em Phys. Rev.} {\bf D 66} (2002) 085016, 
  [\href{http://xxx.lanl.gov/abs/hep-ph/0205085}{{\tt hep-ph/0205085}}].

\bibitem{Andersen:2003zk}
  J.~O.~Andersen, E.~Petitgirard and M.~Strickland, 
  {\it {Two-loop hard-thermal-loop thermodynamics with quarks}},  
  {\em Phys. Rev.} {\bf D 70} (2004) 045001,
  [\href{http://xxx.lanl.gov/abs/hep-ph/0302069}{{\tt hep-ph/0302069}}].

\bibitem{Andersen:2009tc}
  J.~O.~Andersen, M.~Strickland and N.~Su, 
  {\it {Gluon Thermodynamics at Intermediate Coupling}},  
  {\em Phys. Rev. Lett.} {\bf 104} (2010) 122003,
  [\href{http://xxx.lanl.gov/abs/0911.0676}{{\tt arXiv:0911.0676}}].

\bibitem{Andersen:2010ct}
  J.~O.~Andersen, M.~Strickland and N.~Su, 
  {\it {Three-loop HTL gluon thermodynamics at intermediate coupling}},  
  {\em JHEP} {\bf 08} (2010) 113,
  [\href{http://xxx.lanl.gov/abs/1005.1603}{{\tt arXiv:1005.1603}}].

\bibitem{Andersen:2010wu}
  J.~O.~Andersen, L.~E.~Leganger, M.~Strickland and N.~Su, 
  {\it {NNLO hard-thermal-loop thermodynamics for QCD}},  
  {\em Phys. Lett.} {\bf B 696} (2011) 468, 
  [\href{http://xxx.lanl.gov/abs/1009.4644}{{\tt arXiv:1009.4644}}].
  
\bibitem{Frenkel:1989br}
  J.~Frenkel and J.~C.~Taylor,
  {\it {High-temperature limit of thermal QCD}},
  {\em Nucl. Phys.} {\bf B 334 } (1990) 199.  

\bibitem{Taylor:1990ia}
  J.~C.~Taylor and S.~M.~H.~Wong,
  {\it {The effective action of hard thermal loops in QCD}},
  {\em Nucl. Phys.} {\bf B 346 } (1990) 115.
  
\bibitem{Efraty:1992pd}
  R.~Efraty and V.~P.~Nair,
  {\it {Chern-Simons theory and the quark-gluon plasma}},
  {\em Phys. Rev.} {\bf D 47} (1993) 5601,
  [\href{http://xxx.lanl.gov/abs/hep-th/9212068}{{\tt hep-th/9212068}}].
  
\bibitem{Jackiw:1993zr}
  R.~Jackiw and V.~P.~Nair,
  {\it {High temperature response functions and the non-Abelian Kubo formula}},
  {\em Phys. Rev.} {\bf D 48} (1993) 4991,
  [\href{http://xxx.lanl.gov/abs/hep-th/9305241}{{\tt hep-th/9305241}}].
  
\bibitem{Jackiw:1993pc}
  R.~Jackiw, Q.~Liu and C.~Lucchesi,
  {\it {Hard thermal loops, static response, and the composite effective action}},
  {\em Phys. Rev.} {\bf D 49} (1994) 6787,
  [\href{http://xxx.lanl.gov/abs/hep-th/9401002}{{\tt hep-th/9401002}}].
  
\bibitem{Gross:1973id}
  D.~J.~Gross and F.~Wilczek,
  {\it {Ultraviolet Behavior of Non-Abelian Gauge Theories}},
  {\em Phys. Rev. Lett.}  {\bf 30} (1973) 1343.
  
\bibitem{Politzer:1973fx}
  H.~D.~Politzer,
  {\it {Reliable Perturbative Results for Strong Interactions?}},
  {\em Phys. Rev. Lett.}  {\bf 30} (1973) 1346.
  
\bibitem{Braaten:2001vr}
  E.~Braaten and E.~Petitgirard,
  {\it {Solution to the 3-loop $\Phi$-derivable approximation for massless scalar thermodynamics}},
  {\em Phys. Rev.}  {\bf D 65} (2002) 085039,
  [\href{http://xxx.lanl.gov/abs/hep-ph/0107118}{{\tt hep-ph/0107118}}].

\bibitem{Linde:1980ts}
  A.~D.~Linde,
  {\it {Infrared problem in the thermodynamics of the Yang-Mills gas}},
  {\em Phys. Lett.}  {\bf B 96} (1980) 289.
  
\bibitem{Gross:1980br}
  D.~J.~Gross, R.~D.~Pisarski and L.~G.~Yaffe,
  {\it {QCD and instantons at finite temperature}},
  {\em Rev. Mod. Phys.}  {\bf 53} (1981) 43.

\bibitem{Rebhan:1993az}
  A.~K.~Rebhan, 
  {\it {Non-Abelian Debye mass at next-to-leading order}},  
  {\em Phys.Rev.} {\bf D 48} (1993) 3967,
  [\href{http://xxx.lanl.gov/abs/hep-ph/9308232}{{\tt hep-ph/9308232}}].

\bibitem{Carrington:2008dw}
  M.~E.~Carrington, A.~Gynther and D.~Pickering, 
  {\it {Fermion mass at next-to-leading order in the hard thermal loop effective theory}},  
  {\em Phys. Rev.} {\bf D 78} (2008) 045018, 
  [\href{http://xxx.lanl.gov/abs/0805.0170}{{\tt arXiv:0805.0170}}].

\bibitem{Laine:2006cp}
  M.~Laine and Y.~Schroder, 
  {\it {Quark mass thresholds in QCD thermodynamics}},
  {\em Phys. Rev.} {\bf D 73} (2006) 085009,
  [\href{http://xxx.lanl.gov/abs/hep-ph/0603048}{{\tt hep-ph/0603048}}].

\bibitem{Aoki:2005vt}
  Y.~Aoki, Z.~Fodor, S.~D. Katz and K.~K.~Szabo, 
  {\it {The equation of state in lattice QCD: with physical quark masses towards the continuum limit}},      
  {\em JHEP} {\bf 01} (2006) 089,
  [\href{http://xxx.lanl.gov/abs/hep-lat/0510084}{{\tt hep-lat/0510084}}].

\bibitem{Cheng:2009zi}
  M.~Cheng {\em et.~al.}, 
  {\it {Equation of state for physical quark masses}},
  {\em Phys. Rev.} {\bf D 81} (2010) 054504,
  [\href{http://xxx.lanl.gov/abs/0911.2215}{{\tt arXiv:0911.2215}}].

\bibitem{Ipp:2003zr}
  A.~Ipp, G.~D.~Moore and A.~Rebhan, 
  {\it {Comment on and erratum to ``Pressure of hot QCD at large $N_f$''}},  
  {\em JHEP} {\bf 01} (2003) 037,
  [\href{http://xxx.lanl.gov/abs/hep-ph/0301057}{{\tt hep-ph/0301057}}].

\bibitem{Gynther:2009qf}
  A.~Gynther, A.~Kurkela and A.~Vuorinen, 
  {\it {$N_f^3 g^6$ term in the pressure of hot QCD}},  
  {\em Phys. Rev.} {\bf D 80} (2009) 096002,
  [\href{http://xxx.lanl.gov/abs/0909.3521}{{\tt arXiv:0909.3521}}].

\bibitem{Su:2011zv}
  N.~Su, 
  {\em {A gauge-invariant reorganization of thermal gauge theory}},
  \newblock doctoral thesis, 
  {Johann Wolfgang Goethe-Universit\"at Frankfurt am Main}, 2010,
  [\href{http://xxx.lanl.gov/abs/1104.3450}{{\tt arXiv:1104.3450}}].

\bibitem{KorthalsAltes:1999xb}
  C.~Korthals-Altes, A.~Kovner and M.~A.~Stephanov, 
  {\it {Spatial 't Hooft loop, hot QCD and $Z_N$ domain walls}},  
  {\em Phys. Lett.} {\bf B 469} (1999) 205, 
  [\href{http://xxx.lanl.gov/abs/hep-ph/9909516}{{\tt hep-ph/9909516}}].
  
\bibitem{KorthalsAltes:2000gs}
  C.~Korthals-Altes and A.~Kovner,
  {\it {Magnetic $Z_N$ symmetry in hot QCD and the spatial Wilson loop}},
  {\em Phys. Rev.} {\bf D 62} (2000) 096008,
  [\href{http://xxx.lanl.gov/abs/hep-ph/0004052}{{\tt hep-ph/0004052}}].

\bibitem{Pisarski:2000eq}
  R.~D.~Pisarski, 
  {\it {Quark-gluon plasma as a condensate of $SU(3)$ Wilson lines}},  
  {\em Phys. Rev.} {\bf D 62} (2000) 111501,
  [\href{http://xxx.lanl.gov/abs/hep-ph/0006205}{{\tt hep-ph/0006205}}].

\bibitem{Vuorinen:2006nz}
  A.~Vuorinen and L.~G.~Yaffe, 
  {\it {$Z(3)$-symmetric effective theory for $SU(3)$ Yang-Mills theory at high temperature}},  
  {\em Phys. Rev.} {\bf D 74} (2006) 025011, 
  [\href{http://xxx.lanl.gov/abs/hep-ph/0604100}{{\tt hep-ph/0604100}}].

\bibitem{deForcrand:2008aw}
  P.~de~Forcrand, A.~Kurkela and A.~Vuorinen, 
  {\it {Center-symmetric effective theory for high-temperature $SU(2)$ Yang-Mills theory}},  
  {\em Phys. Rev.} {\bf D 77} (2008) 125014, 
  [\href{http://xxx.lanl.gov/abs/0801.1566}{{\tt arXiv:0801.1566}}].

\bibitem{Hidaka:2009hs}
  Y.~Hidaka and R.~D.~Pisarski, 
  {\it {Hard thermal loops, to quadratic order, in the background of a spatial 't Hooft loop}},  
  {\em Phys. Rev.} {\bf D 80} (2009) 036004, 
  [\href{http://xxx.lanl.gov/abs/0906.1751}{{\tt arXiv:0906.1751}}].
  
\bibitem{Dumitru:2010mj}
  A.~Dumitru, Y.~Guo, Y.~Hidaka, C.~P.~K.~Altes and R.~D.~Pisarski,
  {\it {How wide is the transition to deconfinement?}},
  {\em Phys. Rev.} {\bf D 83} (2011) 034022,
  [\href{http://xxx.lanl.gov/abs/1011.3820}{{\tt arXiv:1011.3820}}].  

\end{thebibliography}\endgroup
\bibliographystyle{JHEP}

\end{document}